\documentclass[amsmath,12pt,amssymb,preprint,prd,aps]{revtex4-2}

\usepackage{amsmath}
\usepackage{multirow}

\usepackage{amsfonts}
\usepackage{rotating} 
\usepackage{graphicx} 
\usepackage{caption}
\usepackage{subcaption}
\usepackage{multirow}
\usepackage{bm}
\usepackage{hyperref}
\usepackage{cleveref}
\usepackage[toc,page]{appendix}
\usepackage[justification=raggedright,singlelinecheck=false]{caption}
\usepackage{tabularx}
\usepackage[percent]{overpic}
\usepackage{placeins}

\newcommand{\diracslash}[1]{#1\llap{/\kern2pt}}

\newcommand{\be}{\begin{equation}}
	\newcommand{\ee}{\end{equation}}
\newcommand{\bea}{\begin{eqnarray}}
	\newcommand{\eea}{\end{eqnarray}}
\newcommand{\ba}[1]{\begin{array}{#1}}
	\newcommand{\ea}{\end{array}}
\usepackage{caption}
\newcommand{\bt}{\begin{tabular}}
	\newcommand{\et}{\end{tabular}}

\newcommand{\beas}{\begin{eqnarray*}}
	\newcommand{\eeas}{\end{eqnarray*}}

\begin{document}
	
	\title{Space-like Sachs electric and magnetic form factors of the baryons in the asymmetric nuclear medium}$   $
	\author{Ekta Rawat}
\email{ektarawat1505@gmail.com}
\affiliation{Department of Physics, Dr. B. R. Ambedkar National
	Institute of Technology, Jalandhar, 144008, India}

\author{Navpreet Kaur}
\email{knavpreet.hep@gmail.com}
\affiliation{Department of Physics, Dr. B. R. Ambedkar National
	Institute of Technology, Jalandhar, 144008, India}
	
\author{Harleen Dahiya}
\email{dahiyah@nitj.ac.in}
\affiliation{Department of Physics, Dr. B. R. Ambedkar National
	Institute of Technology, Jalandhar, 144008, India}

\author{Arvind Kumar}
\email{kumara@nitj.ac.in}
\affiliation{Department of Physics, Dr. B. R. Ambedkar National
	Institute of Technology, Jalandhar, 144008, India}

\author{Suneel Dutt}
\email{dutts@nitj.ac.in}
\affiliation{Department of Physics, Dr. B. R. Ambedkar National
	Institute of Technology, Jalandhar, 144008, India}

\date{\today}%

	\def\be{\begin{equation}}
		\def\ee{\end{equation}}
	\def\bearr{\begin{eqnarray}}
		\def\eearr{\end{eqnarray}}
	\def\zbf#1{{\bf {#1}}}
	\def\bfm#1{\mbox{\boldmath $#1$}}
	\def\hf{\frac{1}{2}}
	\def\kp{\zbf k+\frac{\zbf q}{2}}
	\def\km{-\zbf k+\frac{\zbf q}{2}}
	\def\hwo{\hat\omega_1}
	\def\hwt{\hat\omega_2}

\begin{abstract}
In the present work, we have studied the space-like baryon Sachs form factors in the isospin asymmetric nuclear medium using the vector meson dominance (VMD) model. The in-medium effects are incorporated through the medium-modified masses of vector mesons which are calculated using the QCD sum rule approach taking density dependent scalar quark and gluon condensates as inputs from chiral SU(3) quark mean field (CQMF) model. The effective magnetic moments of the baryons are also calculated in the CQMF model. In the framework of VMD model, the photon couples to the nucleons through intermediary vector mesons with the same quantum number as that of a photon. This coupling leads to the relation of isoscalar and isovector Dirac and Pauli form factors which are then used to calculate the Sachs electric and magnetic form factors, which provide physically measurable quantities that represent the electric and magnetic distributions of the baryons. The present work aims to study the effects of asymmetric nuclear matter at finite temperature on the Sachs form factors of baryons in the space-like region. The electric and magnetic charge radii have also been calculated for the baryons in free space and dense asymmetric nuclear matter. The results obtained have been compared with other available phenomenological models, lattice simulations, and experimental data. 

	\end{abstract}
	
	\maketitle
\newpage	
	\section{\label{intro}Introduction}\label{sec:first}
One of the core objectives of hadron physics is to understand the fundamental constituents of matter and the interactions among them. In this context, the internal structure of the nucleons and other baryons remains a subject of fundamental significance. Form factors of the baryons \cite{Hammer:2001rh,Kubon:2001rj,Hyde:2004gef,JeffersonLabE93-026:2003tty,Punjabi:2015bba,Lin:2021xrc,Vaziri:2025rfq,Ramalho:2021inn} provide valuable insights into the internal distributions of quarks and gluons, and serve as a probe of its structure responsible for generating their electromagnetic properties. The deep inelastic scattering experiments performed by the MIT-SLAC Collaboration provided strong evidence for the existence of quarks inside nucleons \cite{Janssens:1965kd}. A historical account of the discovery of quarks is presented in Ref.~\cite{Riordan:1992hr}. While observing EMC effect, which refers to the modification of the parton distributions of nucleons bound in nuclei, the nucleon structure functions deviate from those of free nucleons \cite{Seely:2009gt}. The electromagnetic form factors (EMFFs) of the nucleons offer a direct probe to their internal structure and the corresponding modifications inside the nuclear environment. Despite being neutral, Sachs form factors of the neutrons reveal asymmetric quark distributions, with a negative value of the charge radii and magnetic form factors providing key insights into strong-interaction dynamics \cite{Atac:2021wqj}. \\
Sachs form factors in the space-like region ($Q^2>0$) are measured in elastic electron–nucleon scattering experiments. The experimental determination of space-like form factors requires specialized electron-scattering facilities as the measurements become increasingly difficult at high momentum transfer due to very small scattering cross sections.  In the time-like region ($Q^2<0$), Sachs form factors are studied through the creation or annihilation of nucleon–antinucleon pairs, such as in $e^+e^- \to N\bar{N}$ and $\bar{p}p \to e^+e^-$ processes. Time-like form factors, on the other hand, are investigated through hadron production processes in the time-like region
\cite{Xia:2021agf}. These observables connect experimental measurements with theoretical descriptions of hadronic structure across different physical environments. The vacuum EMFFs for both time-like and space-like regions for the nucleons have been studied widely and act as input for the precision tests of the standard model to determine fundamental constants. They are of critical importance for the accelerator neutrino program.\\

Form factors of baryons have been studied widely in both the space-like and time-like regions, using a large range of theoretical frameworks. These include first-principles approaches such as lattice QCD, where axial form factors have been computed using various lattice formulations and averaging procedures, as well as through the Feynman–Hellmann theorem and domain-wall fermions on highly improved staggered quark (HISQ) ensembles \cite{Sasaki:2007gw}. Parallel approaches involve generalized nonlocal chiral effective theory, dispersively modified chiral perturbation theory, and models based on Skyrme model \cite{Alvarado:2023loi,Kim:2012ts}. Besides this, relativistic descriptions such as the Poincaré-covariant quark–diquark framework employing Faddeev equations have been widely used \cite{Eichmann:2011vu}.
\\ Form factors have also been extracted from generalized parton distributions (GPDs), including recent parametrizations (e.g. GSAMA24), and analysed through their t-dependence in parton distribution functions, short-range correlations, and the EMC effect \cite{Guidal:2004nd,Ghasemzadeh:2025qje,Kim:2024wne}. Related studies include tensor form factors at large $N_e$, electric dipole form factors in the QCD vacuum, and calculations based on extended vector meson dominance (VMD), pion cloud effects, and hard collinear factorization for gravitational form factors \cite{Liu:2025kuc,Wang:2024abv,Yao:2024ixu}. Further results have been obtained from nucleon resonance transition form factors, toy models explaining oscillatory behavior in the time-like region, diagrammatic analyses involving bubble and tadpole contributions, and holographic approaches such as the hard-wall AdS/QCD model \cite{Mamedov:2022tth, Zolfagharpour:2020acg,Gao:2022osh,Ataev:2022tkg}.\\
Experimental facilities such as the relativistic heavy ion collider (RHIC) and the large hadron collider (LHC) probe strongly interacting matter at high temperature and nearly zero baryon density. On the other hand, experiments such as the facility for antiproton and ion research (FAIR), the J-PARC facility at KEK, and NICA at Dubna are dedicated to exploring the high baryon-density regime \cite{Sissakian:2009zza,Kekelidze:2012zzb}. From a theoretical perspective, hadronic models such as quantum hadrodynamics (QHD), the quark-meson coupling (QMC) model \cite{Ramalho:2025kii}, the Nambu-Jona-Lasinio (NJL) model \cite{Carrillo-Serrano:2016igi}, lattice QCD \cite{Lin:2008mr}, vector meson dominance (VMD) model, chiral effective field theory \cite{Yang:2020rpi}, covariant quark-diquark models \cite{Keiner:1995bu}, quark-diquark Faddeev approaches \cite{Liu:2023reo}, and light-front constituent quark models \cite{deAraujo:2022teh}, have been widely employed to investigate medium-induced modifications of hadron properties at finite temperature and density.\\

  The study of the hyperons, which contain strange quarks, is of vital importance as their EMFFs provide additional insights into non-perturbative QCD effects \cite{Yang:2019mzq,Ramalho:2019koj,Lin:2008mr,Yang:2020rpi,BESIII:2020uqk,Haidenbauer:2020wyp} compared to what has been learnt from the nucleons. In the time-like region, EMFFs of the $\Sigma$ hyperon have been studied by BESIII Collaboration \cite{BESIII:2020uqk}. In addition to this, the hyperons EMFFs have been calculated within the lattice QCD \cite{Lin:2008mr}, light cone sum rule \cite{Liu:2009mb}, and chiral perturbative theory (ChPT) \cite{Kubis:2000aa}. Form factors of hyperons such as the $\Lambda$ baryon have been relatively less explored, notably in the space-like region, primarily due to experimental limitations associated with their short lifetimes. Most of the available studies focus on the time-like region, including dispersion-theoretical analyses of EMFFs, VMD-based transition form factors, and investigations at large momentum transfer.  Overall, while nucleon form factors are well constrained across both kinematic regions, hyperon form factors—especially in the space-like domain remain comparatively underexplored and are largely theory-driven.\\
  
The VMD \cite{Schildknecht:2005xr} model provides an effective phenomenological framework in which the photon couples to a baryon through intermediate vector mesons. The resulting Sachs form factors encode the dressing of both the photon-vector-meson and vector-meson-baryon vertices. In an effective sense, they incorporate contributions arising from the internal quark-gluon structure, meson clouds, and other higher-order strong-interaction effects. Based on the recent measurements of the $e^+e^- \rightarrow \Sigma^+\bar{\Sigma}^-$ and $e^+e^- \rightarrow \Sigma^-\bar{\Sigma}^+$, processes by BES III collaboration the parameters of the VMD model has been determined by fitting them to the experimental data on the time-like effective form factors ($\left| G_{\mathrm{eff}} \right|$). In the current work, for $\Sigma^+$ and $\Sigma^-$ hyperon, we have included the contributions from the $\rho$, $\omega$, and $\phi$ mesons to account for the photon–vector meson coupling effects. Within this work, we have also studied the Sachs form factors of $\Lambda$ hyperon , where  we have taken into account the contributions from the resonance states $\omega(1420)$, $\omega(1650)$, $\phi(1680)$ and $\phi(2170)$ \cite{Yang:2019mzq}.  \\
 The $\Lambda$ hyperon, being electrically neutral, has a small electric form factor (EFF) and highly sensitive internal dynamics. Resonance states appear naturally in the VMD diagram via $\gamma^{*} \;\rightarrow\; V(\rho, \omega, \phi) \;\rightarrow\; \Lambda^{*} \;\rightarrow\; \Lambda$. $\Lambda$ hyperon contain a $s-$quark and hence $\phi$ meson couples strongly to these hyperons thus making the resonances quantitatively significant at low and intermediate $Q^2$ behaviour. The inclusion of these resonance states, together with the ground-state contributions, thus provides a more accurate and improved description of the form-factor behaviour. It also gives better non-perturbative QCD effects, which are absent in the pure ground state model. Besides, in nuclear medium, vector meson masses \cite{Hatsuda:1996xt} and widths are modified, which enhances the resonance contributions and make their effect more pronounced. \\
The medium modifications of light hadrons are closely connected to the spontaneous and explicit breaking of chiral symmetry and its restoration at high temperatures and baryon densities. The present study aims to bridge these approaches to provide a coherent description of the electromagnetic structure of  baryons. The current work of the in-medium meson masses and magnetic moments of the baryons provides phenomenological input and acts as a bridge between non-perturbative QCD and experimental data from the large hadron collider beauty experiment (LHCb) \cite{Belyaev:2021cyr} and the future electron-ion collider (EIC) \cite{AbdulKhalek:2022hcn}. 
    In this context, the present work aims to address these existing gaps by extending these models to investigate the internal dynamics of the hadrons in a dense nuclear environment. \\
    
    In the current study, we have focused on the space-like Sachs form factors of the nucleons \cite{Bijker:2005cd} as well as hyperons within an asymmetric nuclear medium at finite temperatures and their comparison with Sachs form factors in free space. These form factors are evaluated using the VMD model, where in-medium effects are incorporated by accounting for the modifications in the masses of light vector mesons as well as the effective magnetic moments of the baryons. The effective masses of vector mesons are determined through the QCD sum rule approach \cite{Mishra:2014rha}, while the effective magnetic moments are calculated in the CQMF model \cite{Singh:2020nwp,Puhan:2024xdq}. The CQMF model is an effective approach in which baryons are described as bound states of constituent quarks interacting with scalar and vector meson fields in a mean-field approximation, and hence the key aspects of QCD and chiral symmetry breaking are incorporated. Within this framework, quarks satisfy the Dirac equation in the presence of background meson fields, such as the scalar-isoscalar field $\sigma$, the scalar-strange field $\zeta$, the vector-isoscalar field $\omega$, and the vector-isovector field $\rho$. The chiral condensates, as $\langle \bar{q}q \rangle$ and $\langle \bar{s}s \rangle$, are directly related to the scalar fields $\sigma$ and $\zeta$, which act as order parameters of chiral symmetry. The values of these condensates are obtained by constructing the thermodynamic potential and thereby, minimizing it with respect to these scalar fields. Solving the coupled nonlinear equations, the solutions provide the density and temperature dependence of the condensates, thereby reflecting the partial restoration of chiral symmetry in hot and dense matter. The in-medium masses are obtained from the QCD sum rules and within VMD framework the vacuum and in-medium form factors are calculated, which are then used to evaluate the root mean square electric and magnetic charge radii. These electric and magnetic charge radii values at vacuum has been compared with available experimental and theoretical results.\\

The present work has been organized as follows: In Sec.~\ref{sec:second}, model framework, used to calculate fields, condensates and the in-medium masses of the light vector mesons, as well as the framework used to calculate the Dirac and Pauli form factors and hence, the Sachs form factors of the baryons have been detailed. In Sec.~\ref{sec:third}, the results and discussions of the Sachs form factors have been presented, and Sec.~\ref{sec:fourth} summarizes the important findings of the present work.

	\section{Methodology} \label{sec:second}

In this section, the theoretical framework for studying the effect of the asymmetric nuclear medium at finite temperature on the Sachs form factors of nucleons and hyperons is presented. In subsection A, we have discussed the CQMF model for the calculations of in-medium scalar quark and gluon condensates. In subsection B, the QCD sum rule to calculate the effective masses of the light vector mesons has been presented. Finally, in subsection C, the VMD model is described for calculating the effective form factors of the baryons.
\vspace{-.6cm}
\subsection{CHIRAL SU(3) QUARK MEAN FIELD MODEL}
\vspace{-.4cm}
In this section, we describe the CQMF model, which has been used to study the in-medium properties of the condensates. The model is based on the spontaneous and explicit breaking of chiral symmetry and also takes into account low energy QCD properties \cite{Papazoglou:1998vr,Weinberg:1968de}. In this model, through a confining potential, quarks are bound inside the baryons and interact via the exchange of scalar fields $\sigma$ and $\zeta$. The contribution of scalar-isovector $\delta$ field becomes significant when the medium has a finite isospin asymmetry. The total effective Lagrangian density in CQMF model is written as \cite{Wang:2001jw}
\vspace{-0.4cm}
	\begin{equation}
		{\cal L}_{{\rm eff}} \, = \, {\cal L}_{q0} \, + \, {\cal L}_{qm}
		\, + \,
		{\cal L}_{\Sigma\Sigma} \,+\, {\cal L}_{vv} \,+\, {\cal L}_{\chi sb}\,  + \, {\cal L}_{C}. \label{totallag}
	\end{equation}
	\label{eq:totallag}
	The first term,  ${\cal L}_{q0} =\bar q \, i\gamma^\mu \partial_\mu \, q$ is the quarks kinetic term. The second term, ${\cal L}_{qm}$ describes the interactions of constituent quarks with the scalar and vector mesons \cite{Wang:2001jw,Singh:2020nwp}. The third term ${\cal L}_{\Sigma\Sigma}$ is the self-interactions of scalar mesons ($\sigma$, $\zeta$ and  $\delta$) and the dilaton field $\chi$ \cite{Wang:2001jw,Singh:2017mxj}. The fourth term gives the self-interactions of vector mesons $\omega$ and $\rho$, and the term ${\cal L}_{\chi sb}$  represents the explicit symmetry breaking. The last term corresponds to the confinement of quarks inside the baryons and is defined as
	\begin{align}
		{\cal L}_{C} = -  \bar \Psi \chi_c \Psi.
	\end{align}
	Here, $ 
	\chi_{c}(r)=\frac14 k_{c} \, r^2(1+\gamma^0)$ is the confining potential, where the coupling constant $k_c$ is taken $98 \ \text{MeV} \text{fm}^{-2}$ \cite{Wang:2001jw}. To study the properties of isospin asymmetric nuclear matter at finite temperature and baryonic density, we have used the mean field approximation. The Dirac equation, under the influence of meson mean fields, for the quark field $\Psi_{qj}$, is given as 
	\cite{Wang:2001jw}
	\begin{equation}
		\left[-i\vec{\alpha}\cdot\vec{\nabla}+\chi_c(r)+\beta m_q^*\right]
		\Psi_{qj}=e_q^*\Psi_{qj}. \label{Dirac}
	\end{equation}

 \hspace{-0.55cm}Here, subscripts $q$ and $j$ denote the quark $q$ ($q$=$u$, $d$) in a nucleon of type $j$ ($j$ = $p$, $n$) in the nuclear medium and $m_q^*$  is the effective quark mass. In terms of the third component of isospin of the respective quark, $I^{3q}$, the effective energy of a particular quark under the influence of the meson field can be expressed as
\begin{equation} 
	e_q^*=e_q-g_\omega^q\omega-g_\rho^q I^{3q}\rho\,.
	\label{eq_eff_energy1}
\end{equation}

\hspace{-0.55cm}Within the CQMF model, the effective mass $\mathcal{M}_j^{*}$ of the $j^{th}$ baryon is related to the effective energy $\mathcal{E}_j^{*}$ and in-medium spurious center of mass momentum $P_{j \, \text{cm}}^{*}$ \cite{Barik:1985rm, Barik:2013lna} through
\begin{align}
	\mathcal{M}_j^*=\sqrt{\mathcal{E}_j^{*2}- \langle P_{j \, \text{cm}}^{*2} \rangle}\,. \label{baryonmass}
\end{align} 
The effective energy $\mathcal{E}_j^{*}$, of baryons can also be expressed in terms of the effective energy of constituent quarks $e_q^*$ using relation 
\begin{equation} 
	\mathcal{E}_j^*=\sum_qn_{qj}e_q^*+\mathcal{E}_{j \, \text{spin}}\,.
	\label{energy}
\end{equation}
In the above equation, 
$n_{qj}$ denotes the number of quarks of type $q$ present in the $j^{th}$ nucleon of medium.
Further, the term $\mathcal{E}_{j \, \text{spin}}$ contributes as a correction to nucleon energy due to spin-spin interaction, which is fitted to the vacuum masses of baryons.

The thermodynamic potential for the isospin asymmetric nuclear matter at finite temperature is 
\begin{equation}
	\begin{split}
		\Omega = -\frac{k_{B}T}{(2\pi)^3} \sum_{j} \alpha_j
		\int_0^\infty d^3k' \biggl\{ {\rm ln}
		\left( 1+e^{- [ \mathcal{E}^{\ast}_j(k') - \beta_j^* ]/k_{B}T}\right)
		+ {\rm ln}\left( 1+e^{- [\mathcal{E}^{\ast}_j(k')+\beta_j^* ]/k_{B}T}\right)
		\biggr\}-{\cal L}_{M}-{\cal V}_{\text{vac}},
	\end{split}
	\label{Eq_therm_pot1}
\end{equation}
where the summation is over the nucleons ($j = p,n$), $\alpha_j=2$ is the degeneracy factor and $\beta_j^*$ is the effective chemical potential. Also, $
{\cal L}_{M} \, = 
{\cal L}_{\Sigma\Sigma} \,+\, {\cal L}_{vv} \,+\, {\cal L}_{\chi sb}\,
$ and $\mathcal{E}^{\ast }(k')=\sqrt{\mathcal{M}_j^{\ast 2}+k'^{2}}$.
The term ${\cal V}_{\text{vac}}$ is subtracted to obtain zero vacuum energy \cite{Wang:2001jw,Puhan:2024xdq}. The thermodynamic potential as mentioned in Eq.~(\ref{Eq_therm_pot1}) is minimised with respect to the scalar fields $\sigma$, $\zeta$, $\delta$, dilaton field $\chi$, and vector fields $\omega$ and $\rho$. Solving the coupled mean-field equations, in-medium meson field expectation values are calculated which are further used to obtain effective constituent quark masses, quark and gluon condensates as well as effective magnetic moments of baryons.

The non-strange quark condensates $\langle u\bar{u} \rangle$ and $\langle d\bar{d} \rangle $ and the strange quark condensate $\langle s\bar{s} \rangle$ are related to the scalar fields $\sigma, \zeta, \delta$ and the scalar dilaton field $\chi$ as

\begin{equation}
	m_u\langle \bar u u \rangle 
	= \frac{1}{2}m_{\pi}^{2} f_{\pi} (\sigma+\delta),       
	\label{eq:eq2}
\end{equation}
\begin{equation}
	m_d \langle \bar d d \rangle
	= \frac{1}{2}m_{\pi}^{2} f_{\pi} (\sigma-\delta),
\label{eq:eq3}
\end{equation}

\begin{equation}
	m_s\langle \bar s s \rangle 
	= \Big( \sqrt {2} m_{k}^{2}f_{k} - \frac {1}{\sqrt {2}} 
	m_{\pi}^{2} f_{\pi} \Big) \zeta,
	\label{eq:eq4}
\end{equation}

\begin{equation}
	\left\langle  \frac{\alpha_{s}}{\pi} {G^a}_{\mu\nu} {G^a}^{\mu\nu} 
	\right\rangle =  \frac{8}{9} \Bigg [(1 - d) \chi^{4}
	+ \left( m_{\pi}^{2} f_{\pi} \sigma
	+ \Big( \sqrt {2} m_{k}^{2}f_{k} - \frac {1}{\sqrt {2}} 
	m_{\pi}^{2} f_{\pi} \Big) \zeta \right) \Bigg ]. 
	\label{eq:eq5}
\end{equation}
The values of the constants are $m_{\pi} = 139$ MeV, $f_{\pi} = 92.8$ MeV, $m_k = 494$ MeV and $f_k = 115$ MeV. For massless quarks, the scalar gluon condensate is proportional to the fourth power of the dilaton field $\chi$, but for finite mass, modification arises from the scalar fields $\sigma$ and $\zeta$. The condensates have been used to calculate the medium-dependent masses of the light vector mesons  $\omega$, $\rho$ and $\phi$ using the QCD sum rule approach.
\subsection{QCD SUM RULES APPROACH}
In the present section, the properties of the vector mesons are studied using QCD sum rules in the nuclear medium \cite{Zschocke:2002mn}. 
The current current correlation function for the vector mesons in terms of the time ordered product is written as
\begin{equation}
	\Pi _{\mu \nu}= i\int d^4x d^4y \langle 0| T j^v_\mu (x) j^v _\nu (0)|0\rangle,
	\label{eq:eq6}
\end{equation}
where $T$ is the time ordered product and $j^v_\mu$ is the 
current for the vector mesons,\\
$j_\mu ^{\rho}=\frac{1}{2} 
(\bar u \gamma_\mu u -\bar d \gamma_\mu d)$,
$j_\mu ^{\omega}=\frac{1}{6} 
(\bar u \gamma_\mu u +\bar d \gamma_\mu d)$, and
$j_\mu ^{\phi}=-\frac{1}{3} (\bar s \gamma_\mu s)$ \cite{Hatsuda:1992bv,Kwon:2008vq}.
We express the correlation function in the space-like region for the light vector mesons in terms of the operator product expansion (OPE). Each term of the OPE contains the coefficient $c^v_i$ with information of the non-perturbative effects of QCD \cite{Hatsuda:1992bv}. For $\phi$ meson these coefficients are defined as 
\begin{equation}
	c_0 ^{\phi}=1+\frac{\alpha_s (Q^2)}{\pi},\;\;\;\;
	c_1 ^{\phi}=-6 {m_s}^2, 
	\label{eq:eq7}
\end{equation}
\begin{equation}
	c_2 ^{\phi}= \frac {\pi^2}{3}
	\langle \frac {\alpha_s}{\pi} G^{\mu \nu} G_{\mu \nu}
	\rangle + 8\pi^2 \langle m_s \bar s s \rangle,
\label{eq:eq8}
\end{equation}
\begin{eqnarray}
	c_3^\phi  = -8\pi^3 \Bigg [ 2 \langle \alpha_s 
	(\bar s \gamma_\mu \gamma_5 \lambda^a s )^2 \rangle
	+ \frac {4}{9} \langle \alpha_s 
	(\bar s \gamma_\mu \lambda^a s )
	(\sum_{q=u,d,s}\bar q \gamma^\mu \lambda^a q) \rangle \Bigg ].
\label{eq:eq9}
\end{eqnarray}
 For the vector mesons $\rho$ and $\omega$,
 \begin{equation}
 	c_0 ^{(\rho,\omega)}=1+\frac{\alpha_s (Q^2)}{\pi},\;\;\;\;
 	c_1 ^{(\rho,\omega)}=-3 (m_u ^2 +m_d ^2),
 	\label{eq:eq10}
 \end{equation}
 \begin{equation}
 	c_2 ^{(\rho,\omega)}= \frac {\pi^2}{3}
 	\langle \frac {\alpha_s}{\pi} G^{\mu \nu} G_{\mu \nu}
 	\rangle + 4\pi^2 \langle m_u \bar u u +m_d \bar d d \rangle,
 	\label{eq:eq11}
 \end{equation}
 \begin{eqnarray}
 	c_3^{(\rho,\omega)} & =& -4\pi^3 \Big [ \langle \alpha_s 
 	(\bar u \gamma_\mu \gamma_5 \lambda^x u 
 	\mp \bar d \gamma_\mu \gamma_5 \lambda^x d )^2 \rangle
 	\nonumber \\
 	&+& \frac {2}{9} \langle \alpha_s 
 	(\bar u \gamma_\mu \lambda^a u 
 	+ \bar d \gamma_\mu \lambda^x d )
 	(\sum_{q=u,d,s}\bar q \gamma^\mu \lambda^x q) \rangle \Big ].
 	\label{eq:eq12}
 \end{eqnarray}
 In the above equation, $\alpha_S =4\pi /(y \ln (Q^2/{\Lambda_{QCD}}^2))$
 is the running coupling constant, with $\Lambda_{QCD} = 140$ MeV
 and $y=11-(2/3)N_f = 9$. 
 In Eq. (\ref{eq:eq12}), the `$\mp$' sign in the first term
 corresponds to $\rho$ and $\omega$ meson, respectively. Comparing the correlator for the vector meson from the Boreal transformation on the phenomenological side and expanding few terms from the operator product expansion \cite{Klingl:1997kf, Kwon:2008vq}, we get the finite energy sum rules (FESRs) for the vector mesons. Using the finite energy sum rules, factorization method \cite{Shifman:1978by} and quark condensates, the coefficient $c_3$ for $\rho$ and $\omega$ vector meson is given as
 \begin{equation}
 	c_3^{(\rho,\omega)}=
 	-\alpha_s \pi^3\times \frac{448}{81} \kappa_q (\langle \bar u u \rangle^2
 	+ \langle \bar d d \rangle^2),
 	\label{eq:eq13}
 \end{equation}
 where, $\kappa_u \simeq \kappa_d =\kappa_q$.
 For the $\phi$ meson,
 \begin{eqnarray}
 	c_3^{\phi} 
 	&=& -8\pi^3 \times \frac{224}{81} \alpha_s \kappa_s 
 	\langle \bar s s \rangle ^2.
 	\label{eq:eq14}
 \end{eqnarray}
 
  From the Boreal transformation of the spectral functions as discussed earlier, we obtain the finite sum rules for vacuum which are given as \cite{Klingl:1997kf, Kwon:2008vq}
 \begin{equation}
 	H_v =
 	{d_v} (c^v_0 s^v_0 +c^v_1) , 
 	\label{eq:eq15}
 \end{equation}
 \begin{equation}
 	H_v m_v^2=
 	{d_v} \Big (
 	\frac {(s^v_0)^2 c^v_0 }{2}-c^v_2 \Big ),
 	\label{eq:eq16}
 \end{equation}
 \begin{equation}
 	H_v m_v^4=
 	{d_v} \Big (
 	\frac{(s^v_0)^3}{3} c^v_0 +c^v_3 \Big ).
 	\label{eq:eq17}
 \end{equation}
 
 \hspace{-.5cm}Eqs.~\eqref{eq:eq15}-~\eqref{eq:eq17} are fitted to the vacuum masses of the mesons to find the vacuum values of the four quark condensates $c^v_3$ and the value of $k_v$. Here, $ d_v = \frac{3}{2},\ \frac{1}{6},\ \text{and}\ \frac{1}{3}
 	\ \text{for} \ \rho,\ \omega,\ \text{and}\ \phi\ \text{}$
  mesons, respectively. At nonzero baryon densities, the spectral function of vector mesons acquires an additional contribution due to their scattering with baryons present in the medium. Consequently, the Borel-transformed dispersion relation for the vector mesons is modified as 
 \begin{equation}
 	\int_{0}^{\infty} ds\, e^{-s/{M}^{2}}\, R_{{phen}}^{v}(s) + 12\pi^{2}\,\Pi^{v}(0)
 	= d_{v}\!\left[c_{0}^{v} {M}^{2} + c_{1}^{v} + \frac{c_{2}^{v}}{{M}^{2}} + \frac{c_{3}^{v}}{2{M}^{4}}\right],
 	\label{eq:eq18}
 \end{equation}
 where the additional term $\Pi^{v}(0)$ denotes the forward scattering amplitude of the vector meson $v$ in the medium and $M$ is the Boreal mass. In nuclear matter, the forward scattering amplitude can be written as 
 \begin{equation}
 	\Pi^{\omega, \rho}(0) = \frac{\rho_{j}}{4\mathcal M^*_{N}}.
  \hspace{0.3cm}
 	\label{eq:eq19}
 \end{equation}
For the $\phi$ meson it vanishes as the $\phi$–nucleon coupling is negligible.
 The coefficients $c_{0}^{v}$ and $c_{1}^{v}$ in Eq.\,(\ref{eq:eq10}) remain independent of the medium, whereas $c_{2}^{v*}$ and $c_{3}^{v*}$ depend explicitly on the in-medium quark and gluon condensates, as defined by Eqs.~(\ref{eq:eq11})--(\ref{eq:eq14}). 
 The medium modification of the vector meson spectral function arises from the change in both condensates and the additional scattering term $\Pi^{v}(0)$, reflecting the baryonic composition of the system. In the text, $c_2^{v*}$, $c_3^{v*}$ and $s^{v*}_0$ are being used to depict the medium dependent effects. At finite densities, FESR for the vacuum is modified to
 \begin{equation}
 	H^*_v =
 	{d_v} ({c^v_0} {s^{v*}_0} +{c^v_1}) -12\pi^2 \Pi^v(0),
 \label{eq:eq24}	
 \end{equation}
 \begin{equation}
 	H^*_v {m^*_v}^2=
 	{d_v} \Big (
 	\frac {(s^{v*}_0)^2 c^v_0}{2}-c_2^{v*}\Big ),
 \label{eq:eq25}
 \end{equation}
 \begin{equation}
 	H^*_v {m^*_v}^4=
 	{d_v} \Big (
 	\frac{(s^{v*}_0)^3}{3} c^v_0 +c_3^{v*} \Big ).
 	\label{eq:eq26}
 \end{equation}
 
\hspace{-0.55cm}We have simultaneously solved these equations to obtain the values of $m^*_v$, $s^{v*}_0$, $H^*_v$ using the coefficient $k_v$ of the four quark condensate for the vector mesons as determined from the FESRs rules in vacuum.
\vspace{-0.5cm}
\subsection{VECTOR MESON DOMINANCE MODEL} 
\vspace{-0.2cm}
In the present section, we will calculate the EMFFs of the baryons taking in-medium magnetic moments from the CQMF model and in-medium masses of the light vector mesons from the QCD sum rules. 
Relativistic invariance allows the nucleon current to be expressed in terms of the Dirac form factor $F_{1}$ and the Pauli form factor $F_{2}$ through the following relation \cite{Bijker:2005cd}
\vspace{-0.29cm}
\begin{equation}
J^{\mu} =\gamma^{\mu}F_{1}(Q^{2})\ + i\,\frac{\sigma^{\mu\nu}F_{2}(Q^{2}) q_{\nu}}{2 \mathcal M_{b}}\,,
\label{eq:eq27}	
\end{equation}
where $\mathcal M_{b}$ denotes the mass of the baryon, $b = p, n, \Sigma^{\pm}, \Lambda$, $q_{\nu}$ is the four-momentum transfer, and $\sigma^{\mu\nu} = \frac{i}{2}[\gamma^{\mu}, \gamma^{\nu}]$.

In the framework of VMD model, the coupling of photon to the nucleon occurs indirectly through the exchange of light vector mesons \cite{Iachello:1972nu}. These interactions lead to relations connecting the electric and magnetic \textit{Sachs form factors} of the nucleon, $G_{E}$ and $G_{M}$, respectively with the isoscalar $F^{S*}$ and isovector $F^{V*}$  form factors. We have
\vspace{-0.3cm}
\begin{align}
	G^{p*}_{M} &= \left(F_{1}^{S*} + F_{1}^{V*}\right)
	+ \left(F_{2}^{S*} + F_{2}^{V*}\right),
	 \\
	G^{p*}_{E} &= \left(F_{1}^{S*} + F_{1}^{V*}\right)
	- \tau_p \left(F_{2}^{S*} + F_{2}^{V*}\right),
 \\
	G^{n*}_{M} &= \left(F_{1}^{S*} - F_{1}^{V*}\right)
	+ \left(F_{2}^{S*} - F_{2}^{V*}\right),
 \\
	G^{n*}_{E} &= \left(F_{1}^{S*} - F_{1}^{V*}\right)
	- \tau_n \left(F_{2}^{S*} - F_{2}^{V*}\right).
	\label{eq:eq28}
\end{align}

In VMD model \cite{Iachello:1972nu}, the Dirac form factor receives contributions from both the intrinsic three-quark structure and the meson cloud, whereas the Pauli form factor is largely dominated by the meson cloud. Taking into consideration the dimensional counting rules \cite{Brodsky:1973kr, Brodsky:1974vy} and perturbative QCD (p-QCD), the term of the type $g(Q^{2})=1/(1+\gamma Q^{2})^2$ is introduced in $F_{2}^{V*}$ and $\tau_b={Q^2/4\mathcal M^2_b} $. Using these, the expressions for isoscalar and isovector Dirac and Pauli form factors are expressed as \cite{Bijker:2004yu}.
	\begin{align}
		F_{1}^{S*}(Q^{2}) &= \frac{1}{2}g(Q^{2})
		\left[
		1-\beta_{\omega}-\beta_{\varphi}
		+\beta_{\omega}\frac{m_{\omega}^{*2}}{m_{\omega}^{*2}+Q^{2}}
		+\beta_{\varphi}\frac{m_{\varphi}^{*2}}{m_{\varphi}^{*2}+Q^{2}}
		\right], \notag \\[2mm]
		F_{1}^{V*}(Q^{2}) &= \frac{1}{2}g(Q^{2})
		\left[
		1-\beta_{\rho}
		+\beta_{\rho}\frac{m_{\rho}^{*2}}{m_{\rho}^{*2}+Q^{2}}
		\right], \notag \\[2mm]
		F_{2}^{S*}(Q^{2}) &= \frac{1}{2}g(Q^{2})
		\left[
		\left(\mu^*_{p}+\mu^*_{n}-1-\alpha_{\varphi}\right)
		\frac{m_{\omega}^{*2}}{m_{\omega}^{*2}+Q^{2}}
		+\alpha_{\varphi}\frac{m_{\varphi}^{*2}}{m_{\varphi}^{*2}+Q^{2}}
		\right], \notag \\[2mm]
		F_{2}^{V*}(Q^{2}) &= \frac{1}{2}g(Q^{2})
		\left[
		\frac{\mu^*_{p}-\mu^*_{n}-1-\alpha_{\rho}}
		{1+\gamma Q^{2}}
		+\alpha_{\rho}\frac{m_{\rho}^{*2}}{m_{\rho}^{*2}+Q^{2}}
		\right].
		\label{ff2}
	\end{align}
The effective magnetic moments $\mu^*_{p}$ and $\mu^*_{n}$ of the nucleons appearing in the above equation have been calculated within the CQMF model which includes the contribution from valence quark, sea quarks and orbital quarks \cite{Singh:2017mxj,Singh:2016hiw}.

%
 
 The values of the meson masses are taken as $m_\rho = 0.770$ GeV, $m_\omega = 0.783$ GeV, and $m_\phi = 0.783$ GeV while the values of the constants used in the calculations of the Sachs form factors of the nucleons are $\beta_{\rho}=0.512$, 
 $\beta_{\omega}=1.129$, $\beta_{\varphi}=-0.263$, $\alpha_{\rho}=2.675$, 
 $\alpha_{\varphi}=-0.200$ and $\gamma=0.515$ (GeV/c)$^{-2}$ \cite{Bijker:2004yu}. The decay width of the $\rho$ meson is crucial for the small $Q^{2}$ behavior of the form factors and is taken into account through the replacement \cite{Iachello:1972nu,Frazer:1960zzb}
\begin{equation}
	\frac{m^{*2}_{\rho}}{m^{*2}_{\rho}+Q^{2}}
	\rightarrow
	\frac{m^{*2}_{\rho} + \dfrac{8\Gamma^{*}_\rho m^{*}_{\pi}}{\pi}}
	{m^{*2}_{\rho} + Q^{2}
		+ \left(4m^{*2}_{\pi} + Q^{2}\right)
		\dfrac{\Gamma^{*}_\rho\,\alpha(Q^{2})}{m^{*}_{\pi}}},
	\label{eq:eq29}
\end{equation}
with 
 \begin{equation}
 	\alpha \left( Q^{2}\right) =\frac{2}{\pi} \sqrt{\frac{4m^{*2}_{\pi}+Q^2}{Q^2}} 
 	\, \ln \left(\frac{\sqrt{ 4m^{*2}_{\pi }+Q^{2}}+\sqrt{Q^{2}}}{%
 		2m^{*}_{\pi }}\right) ~.  
 		\label{eq:eq30}
 \end{equation}
The in-medium decay width $\Gamma^{*}_\rho$ \cite{Zschiesche:2003qq}, is calculated using
 \begin{equation}
 	\Gamma^{*}_{\rho} = 
 	\frac{g^{2}_{\rho \pi \pi}}{48\pi} 
 	m^{*}_{\rho}
 	\left( 1 - \frac{4m_{\pi}^{2}}{m_{\rho}^{*2}} \right)^{3/2}
 	(X^2 - Y^2).
 	\label{eq:eq31}
 \end{equation}
 
 \hspace{-0.5cm}In the above equation,
 $
 X = \left( 1 + f\left( \frac{m^{*}_{\rho}}{2} \right) \right)
 \quad \text{and} \quad
 Y = f\left( \frac{m^{*}_{\rho}}{2} \right).
$
The function 
  $
 f(\frac{m^{*}_{\rho}}{2}) = \exp(\frac{m_{\rho}^*}{2(T-1)})^{-1}
$ represents the Bose-Einstein distribution function \cite{Kumar:2022hps}. The coupling constant $g_{\rho \pi \pi}$ is fixed to the vacuum value of the decay width $0.112$ GeV of the $\rho$ meson. 
 For $\Sigma^{\pm}$ hyperon, the Dirac and Pauli form factors have been calculated by taking intrinsic form factor $
 g_{1}(Q^{2})
 =
 (1+\gamma_{1}Q^{2})^{-2}
 \quad \text{and} \quad
 g_{2}(Q^{2})
 =
 (1+\gamma_{2}Q^{2})^{-2}
 $  \cite{Li:2020lsb}.\\
 \\
The free model parameters $\gamma_{1}$ and $\gamma_{2}$ as well as other constants used in the calculations of Sachs form factors of $\Sigma^{\pm}$ hyperons are presented in Table~\ref{tab:table1} \cite{Li:2020lsb}. The Sachs form factors for the $\Sigma^{\pm}$ baryon are given as
 \begin{align}
 	G^{\Sigma{^*\pm}}_E(Q^2) &= F^{\Sigma{^*\pm}}_1 - \tau_{\Sigma^{^\pm}} F^{\Sigma{^*\pm}}_2 
 	= F_{1}^{S*} + F_{1}^{V*}
 	- \tau \left(F_{2}^{S*} + F_{2}^{V*}\right), \\
 	G^{\Sigma{^*\pm}}_M(Q^2) &= F^{\Sigma{^*\pm}}_1 + F^{\Sigma{^*\pm}}_2 
 	= F_{1}^{S*} + F_{1}^{V*}
 	+ F_{2}^{S*} + F_{2}^{V*},
 	\label{eq:eq35}
 \end{align}
 where $\tau_{\Sigma^{^\pm}}$ = $\frac{Q^2}{4M^2_\Sigma{^\pm}}$ and $M_\Sigma{^\pm}$ being the vacuum mass of the hyperon. The isoscalar and isovector form factors of $\Sigma{^\pm}$ can now be expressed as \cite{Li:2020lsb}
 
\begin{eqnarray}
	F^{S*}_{1 \Sigma^{+}}\left(Q^{2}\right) &=&
	\frac{1}{2} g_{1}\left(Q^{2}\right)
	\left[
	\left(1-\beta_{\omega}-\beta_{\phi}\right)
	+ \beta_{\omega} \frac{(m_{\omega}^{*})^{2}}{(m_{\omega}^{*})^{2}+Q^{2}}
	+ \beta_{\phi} \frac{(m_{\phi}^{*})^{2}}{(m_{\phi}^{*})^{2}+Q^{2}}
	\right],  \label{eq:f1sigmaps} \\	
	F^{V*}_{1 \Sigma^{+}}\left(Q^{2}\right) &=&
	\frac{1}{2} g_{1}\left(Q^{2}\right)
	\left[
	\left(1-\beta_{\rho}\right)
	+ \beta_{\rho} \frac{(m_{\rho}^{*})^{2}}{(m_{\rho}^{*})^{2}+Q^{2}}
	\right],  \label{eq:f1sigmapv} \\
	F^{S*}_{2 \Sigma^{+}}\left(Q^{2}\right) &=&
	\frac{1}{2} g_{1}\left(Q^{2}\right)
	\left[
	\left(2\mu_{\Sigma^+}^{*} - 2 -\alpha_{\phi}-\alpha_{\rho}\right)
	\frac{(m_{\omega}^{*})^{2}}{(m_{\omega}^{*})^{2}+Q^{2}}
	+ \alpha_{\phi} \frac{(m_{\phi}^{*})^{2}}{(m_{\phi}^{*})^{2}+Q^{2}}
	\right],  \label{eq:f2sigmaps} \\
	F^{V*}_{2 \Sigma^{+}}\left(Q^{2}\right) &=&
	\frac{1}{2} g_{1}\left(Q^{2}\right)
	\left[
	\alpha_{\rho}
	\frac{(m_{\rho}^{*})^{2}}{(m_{\rho}^{*})^{2}+Q^{2}}
	\right],  \label{eq:f2sigmapv} \\
	F^{S*}_{1 \Sigma^{-}}\left(Q^{2}\right) &=&
	\frac{1}{2} g_{2}\left(Q^{2}\right)
	\left[
	\left(-1-\beta_{\omega}-\beta_{\phi}\right)
	+ \beta_{\omega} \frac{(m_{\omega}^{*})^{2}}{(m_{\omega}^{*})^{2}+Q^{2}}
	+ \beta_{\phi} \frac{(m_{\phi}^{*})^{2}}{(m_{\phi}^{*})^{2}+Q^{2}}
	\right],  \label{eq:f1sigmamfs} \\
	F^{V*}_{1 \Sigma^{-}}\left(Q^{2}\right) &=&
	\frac{1}{2} g_{2}\left(Q^{2}\right)
	\left[
	\left(-1-\beta_{\rho}\right)
	+ \beta_{\rho} \frac{(m_{\rho}^{*})^{2}}{(m_{\rho}^{*})^{2}+Q^{2}}
	\right],  \label{eq:f1sigmamfv} \\
	F^{S*}_{2 \Sigma^{-}}\left(Q^{2}\right) &=&
	\frac{1}{2} g_{2}\left(Q^{2}\right)
	\left[
	\left(2\mu_{\Sigma^-}^{*} + 2 -\alpha_{\phi}-\alpha_{\rho}\right)
	\frac{(m_{\omega}^{*})^{2}}{(m_{\omega}^{*})^{2}+Q^{2}}
	+ \alpha_{\phi} \frac{(m_{\phi}^{*})^{2}}{(m_{\phi}^{*})^{2}+Q^{2}}
	\right],  \label{eq:f2sigmamfs} \\
	F^{V*}_{2 \Sigma^{-}}\left(Q^{2}\right) &=&
	\frac{1}{2} g_{2}\left(Q^{2}\right)
	\left[
	\alpha_{\rho}
	\frac{(m_{\rho}^{*})^{2}}{(m_{\rho}^{*})^{2}+Q^{2}}
	\right].  \label{eq:f2sigmamfv}
	\label{eq:eq36}
\end{eqnarray}

\hspace{-0.5cm}The effective magnetic moments $\mu^{*}_{\Sigma{^\pm}}$ gives total magnetic moment of the $\Sigma$ hyperon calculated from CQMF model which includes contribution from valence, sea and orbital quarks.
\begin{table}[htbp]
	\captionsetup{width=1.5\textwidth}	
	\setlength{\tabcolsep}{39pt}
	\begin{tabular}{|c|c|c|c|}
		\hline
		\textrm{Parameter} & \textrm{Value} & \textrm{Parameter} & \textrm{Value} \\
		\hline
		$\beta_{\rho}$   & 0.736  & $\alpha_{\rho}$ & 0.976 \\
		\hline
		$\beta_{\phi}$   & $-$0.441 & $\alpha_{\phi}$ & 1.035 \\
		\hline
		$\beta_{\omega}$ & 0.434  & $\gamma_{1}$    & 0.46  \\
		\hline
		$\gamma_{2}$     & 1.18   &                &       \\
		\hline
	\end{tabular}
	\caption{The parameter values used for the calculation of the $\Sigma{^\pm}$ \\baryon's Sachs form factors \cite{Li:2020lsb}.}
	\label{tab:table1}
\end{table}
\\The EMFF of the $\Lambda$ hyperon gets contribution from the isoscalar channel only and are given as \cite{Yang:2019mzq}
 \begin{align*}
 	G^{\Lambda *}_{M} &= F_{1\Lambda}^{S*} + F_{2\Lambda}^{S*},
 	\quad \text{and} \quad
 	G^{\Lambda *}_{E} = F_{1\Lambda}^{S*} - \tau_\Lambda F_{2\Lambda}^{S*},
 \end{align*}
  where $\tau_\Lambda={Q^2/4\mathcal M^2_\Lambda}$ and $\mathcal M_\Lambda$ denotes vacuum mass of the $\Lambda$ hyperon.
Taking into account the isoscalar property of $\Lambda$ hyperon, the contribution of $\rho$ meson and its resonance states has been excluded. In this work, we have included the contributions from the resonance states \cite{Yang:2019mzq}.
 Considering p-QCD and asymptotic behaviour of the form factors and including the constraints leading to parameterised forms \cite{Brodsky:2003gs,Belitsky:2002kj}, we get the isoscalar Dirac and Pauli form factors for $\Lambda$ hyperon which are respectively expressed as
 \vspace{-0.8cm}
 \begin{eqnarray*}
 	F_{1\Lambda}^{S*}(Q^{2}) &=&
 	\frac{g(Q^{2})}{3}
 	\sum_{i=1}^{I}
 	\bigg[
 	-b_{\omega_i}-b_{\phi_i}
 	+ b_{\omega_i}
 	\frac{(m_{\omega_i}^{*})^{2}}
 	{(m_{\omega_i}^{*})^{2}+Q^{2}}
 	+ b_{\phi_i}
 	\frac{(m_{\phi_i}^{*})^{2}}
 	{(m_{\phi_i}^{*})^{2}+Q^{2}}
 	\bigg],
 	\label{F1s} \\ 
 	\quad \text{} \quad\\
 	F_{2\Lambda}^{S*}(Q^{2}) &=&
 	\frac{g(Q^{2})}{3}
 	\sum_{i=1}^{I}
 	\bigg[
 	(\mu^*_{\Lambda}-a_{\phi_i})
 	\frac{(m_{\omega_i}^{*})^{2}}
 	{(m_{\omega_i}^{*})^{2}+Q^{2}}
 	+ a_{\phi_i}
 	\frac{(m_{\phi_i}^{*})^{2}}
 	{(m_{\phi_i}^{*})^{2}+Q^{2}}
 	\bigg].
 	\label{eq:eq38}
 	\end{eqnarray*}

\hspace{-0.5cm}Here, $I=3$ represents the number of resonance states, where 
$\omega_i \ (i=1,2,3)$ correspond to the vector mesons $\omega(782)$, $\omega(1420)$, and $\omega(1650)$, while $\phi_i$ denote the vector mesons $\phi(1020)$, $\phi(1680)$, and $\phi(2170)$. The values of free parameters and coefficients used in the current study of the calculation of the Sachs form factor of the $\Lambda$ hyperon are listed in Tables~\ref{Tab:table2} and~\ref{tab:table3}.
 
\begin{table}[htb]
	\centering
\captionsetup{width=0.83\textwidth}	
\setlength{\tabcolsep}{10pt}   
\begin{tabular}{|c|c|c|c|c|c|}
	\hline
		\toprule
		\text{State} & \text{Mass} & \text{Width} & \text{State} & \text{Mass} & \text{Width} \\
		\toprule
		$\omega(782)$   & 782  & 8.1  & $\phi(1020)$  & 1019 & 4.2  \\ \hline
		$\omega(1420)$  & 1418 & 104  & $\phi(1680)$  & 1674 & 165  \\ \hline
		$\omega(1650)$  & 1679 & 121  & $\phi(2170)$  & 2171 & 128   \\ \hline
		
	\end{tabular}
\hspace{1.4cm}	\caption{ Masses and widths of the vector mesons 
	 (in MeV) \\used in the calculation of the Sachs form factor of $\Lambda$ hyperon \cite{Yang:2019mzq}.}
	\label{Tab:table2}
\end{table}
\setlength{\tabcolsep}{10pt}   
\begin{table}[htb]
\captionsetup{width=0.85\textwidth}		
\begin{tabular}{|c|c|c|c|c|c|}
		\toprule
		Parameter & Value & Parameter &Value & Parameter &Value \\
		\toprule 
		$b_{\omega(782)}$ & $1.248$  &
		$b_{\omega(1420)}$ & $0.712$  &
		$b_{\omega(1650)}$ & $1.013$ \\ \hline
		$b_{\phi(1020)}$ &  $-1.902$ &
		$b_{\phi(1680)}$ & $-0.581$  &
		$b_{\phi(2170)}$ & $ -0.584$\\ \hline
		$a_{\phi(1020)}$ & $-2.224$  &
		$a_{\phi(1680)}$ & $2.748$  &
		$a_{\phi(2170)}$ & $0.615$ \\ \hline
		
	\end{tabular}
\hspace{0.45cm}	\caption{The parameter values used for the calculation of the Sachs form factors of $\Lambda$ hyperon \cite{Yang:2019mzq,Iachello:1972nu}. \label{tab:table3}}
\end{table}

\begin{table}[htbp]
	\centering
	\begin{tabular}{|c|c|c|c|c|c|c|}
		\hline
		S.No. & Charge radii & Experimental Value & VMD Model & \multicolumn{3}{c|}{Current work} \\
		\cline{5-7}
		&  &  &  & $\rho_B=0$ & $\rho_B=\rho_0/2$ & $\rho_B=\rho_0$ \\
		\hline
		
		1 & $\langle r_E^{p\,2} \rangle^{1/2}$ 
		& 0.862 $\pm 0.012$\cite{Simon:1980hu} 
		& 0.838 \cite{Bijker:2005cd} 
		& 0.829 & 0.8698 & 0.9217 \\
		\hline
		
		& $\langle r_M^{p\,2} \rangle^{1/2}$ 
		& 0.855 $\pm 0.035$\cite{Hyde:2004gef} 
		& 0.825 \cite{Bijker:2005cd} 
		& 0.835 & 0.8517 & 0.8726 \\
		\hline
		
		2 & $\langle r_E^{n\,2} \rangle$ 
		& -0.115 $\pm 0.003$\cite{Kopecky:1997rw} 
		& -0.100 \cite{Bijker:2005cd} 
		& -0.098 & -0.0907 & -0.0778 \\
		\hline
		
		& $\langle r_M^{n\,2} \rangle^{1/2}$ 
		& 0.873 $\pm 0.011$\cite{Kubon:2001rj} 
		& 0.834 \cite{Bijker:2005cd} 
		& 0.844 & 0.8487 & 0.8522 \\
		\hline
		
		3 & $\langle r_E^{\Lambda\,2} \rangle^{1/2}$ 
		& - 
		& 0.11 \cite{Yang:2019mzq} 
		& 0.154 & 0.1844 & 0.2211 \\
		\hline
		
		& $\langle r_M^{\Lambda\,2} \rangle^{1/2}$ 
		& - 
		& 0.42 \cite{Yang:2019mzq} 
		& 0.294 & 0.3167 & 0.3462 \\
		\hline
		
		4 & $\langle r_E^{\Sigma^+\,2} \rangle^{1/2}$ 
		& - 
		& - 
		& 0.561 & 0.6042 & 0.6616 \\
		\hline
		
		& $\langle r_M^{\Sigma^+\,2} \rangle^{1/2}$ 
		& - 
		& - 
		& 0.587 & 0.6307 & 0.6910 \\
		\hline
		
		5 & $\langle r_E^{\Sigma^-\,2} \rangle$ 
		& - 
		& - 
		& -0.275 & -0.2375 & -0.1865 \\
		\hline
		
		& $\langle r_M^{\Sigma^-\,2} \rangle^{1/2}$ 
		& - 
		& - 
		& 0.359 & 0.3862 & 0.4281 \\
		\hline
		
	\end{tabular}
	
	\caption{ The electric and magnetic charge radii of the baryons are compared with VMD model and available experimental data. The quantities $\langle r_E^{n\,2} \rangle$ and $\langle r_E^{\Sigma^-\,2} \rangle$ are in $\mathrm{fm}^2$, whereas for other baryons the results are in fm.}
	\label{tab:table4}
\end{table}

	\begin{figure}[b]
	\includegraphics[height=0.44\textwidth,width=1.1\textwidth]{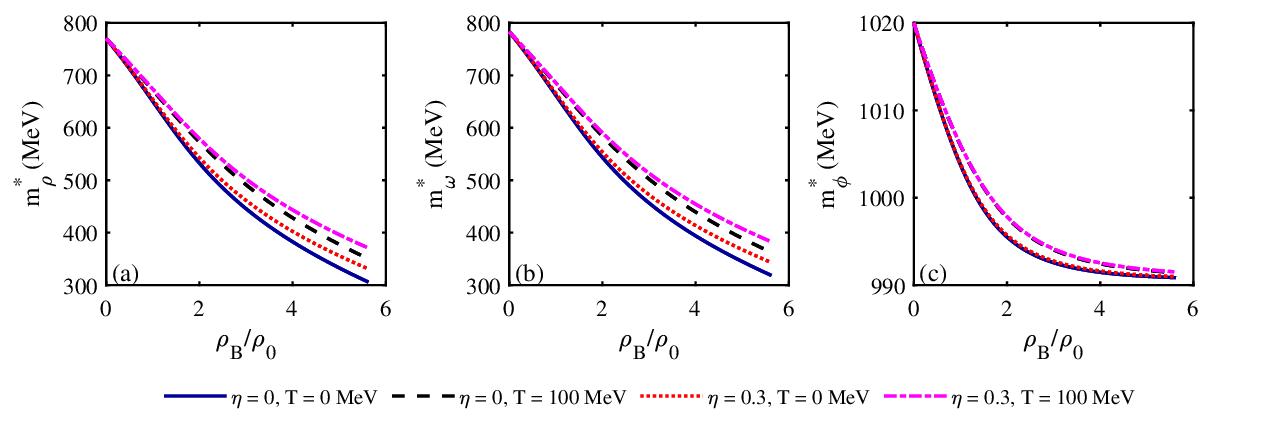}
	\caption{Variation in the masses of the light vector mesons (a) $\rho$, (b) $\omega$, (c) $\phi$ as a function of baryonic density $\rho_B$ (in the units of $\rho_0$) at $T=0$ and $100$ MeV as well as with isospin asymmetry, $\eta= 0$ and $0.3$ .}
	\label{fig:fig0}
\end{figure}
		
\begin{figure}[h!]
	
		\includegraphics[height=1.0\textwidth,width=1.0\textwidth]{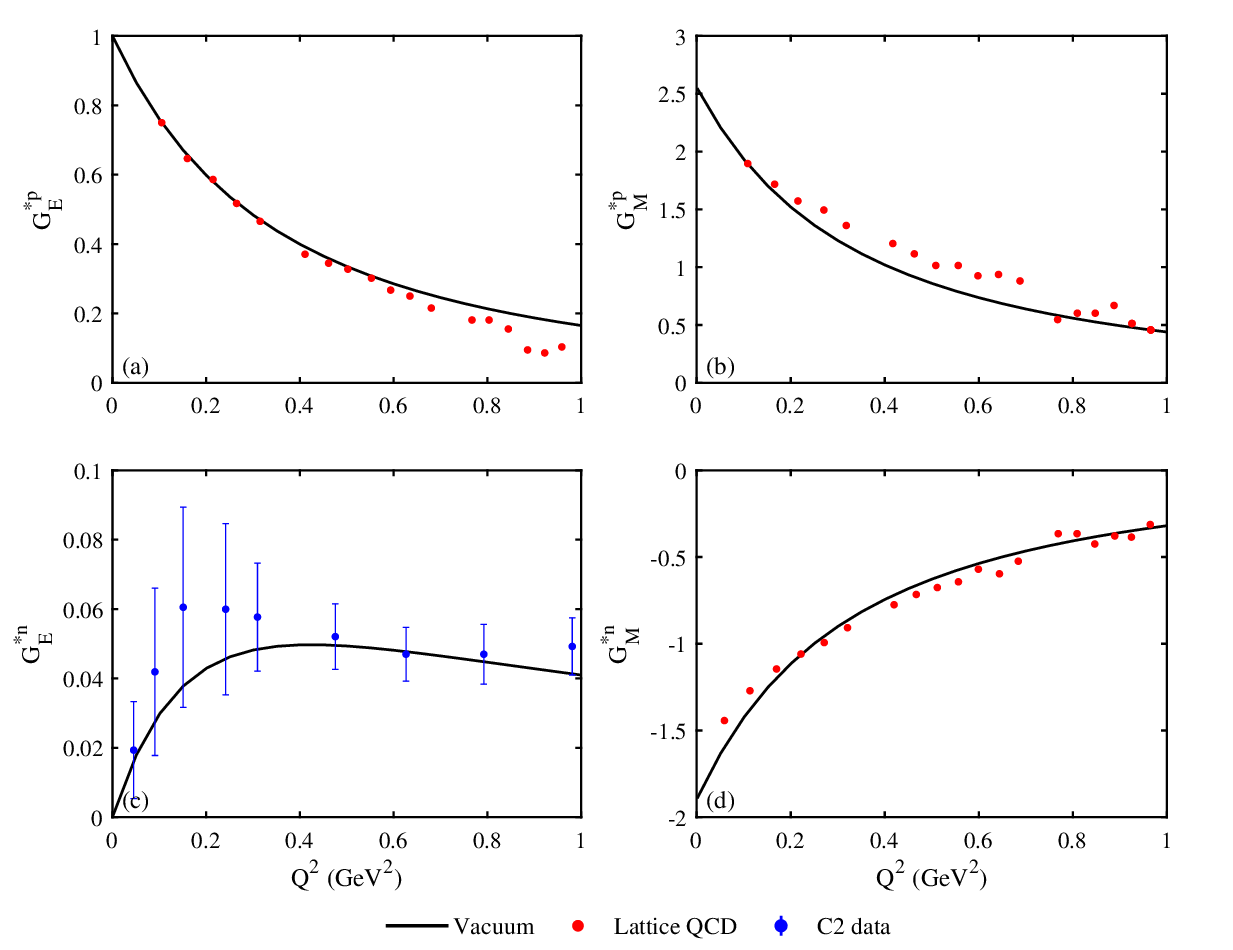}
		
	\caption{ The vacuum ($\rho_B=0$, $\eta=0$, $T=0$ MeV) Sachs electric form factor $ G^{*p,n}_{E} $ (subplots (a) and (c)) and Sachs magnetic form factor $ G^{*p,n}_{M} $ (subplots (b) and (d)) of the baryons $p, n$ are shown as a function of momentum transfer $Q^2$ (in GeV$^2$). Comparison has been made with the  available experimental results for $G^{*n}_{E} $ \cite{Schiavilla:2001qe} and lattice QCD results for $G^{*p}_{E} $ and  $ G^{*p,n}_{M} $ \cite{Alexandrou:2018sjm}.}
	\label{fig:figure01}
	
\end{figure}
	
		\begin{figure}[h!]
		\includegraphics[height=0.85\textwidth,width=1.1\textwidth]{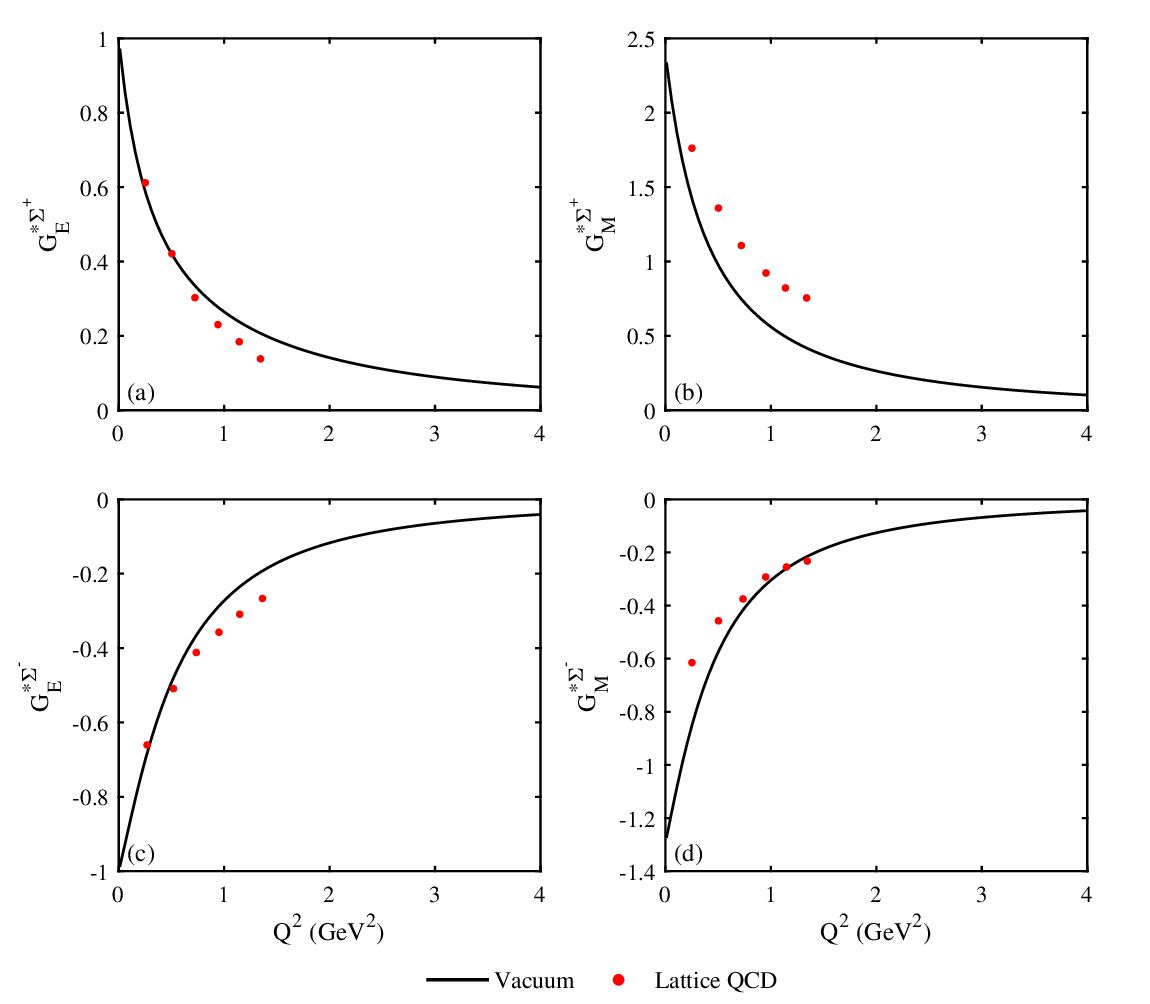}

		\caption{The vacuum ($\rho_B=0$, $\eta=0$, $T=0$ MeV) Sachs electric form factor $G^{*\Sigma^{\pm}}_{E}$ (subplots (a) and (c)) and Sachs magnetic form factor $G^{*\Sigma^{\pm}}_{M}$ (subplots (b) and (d)) of the $\Sigma^{\pm}$ hyperons are shown as a function of momentum transfer $Q^2$ (in GeV$^2$). Comparison has been made with the available lattice QCD results \cite{Li:2020lsb}.}
		\label{fig:figure02}
	\end{figure}
		
	\begin{figure}[h!]
				
			\includegraphics[height=1.0\textwidth,width=1.0\textwidth]{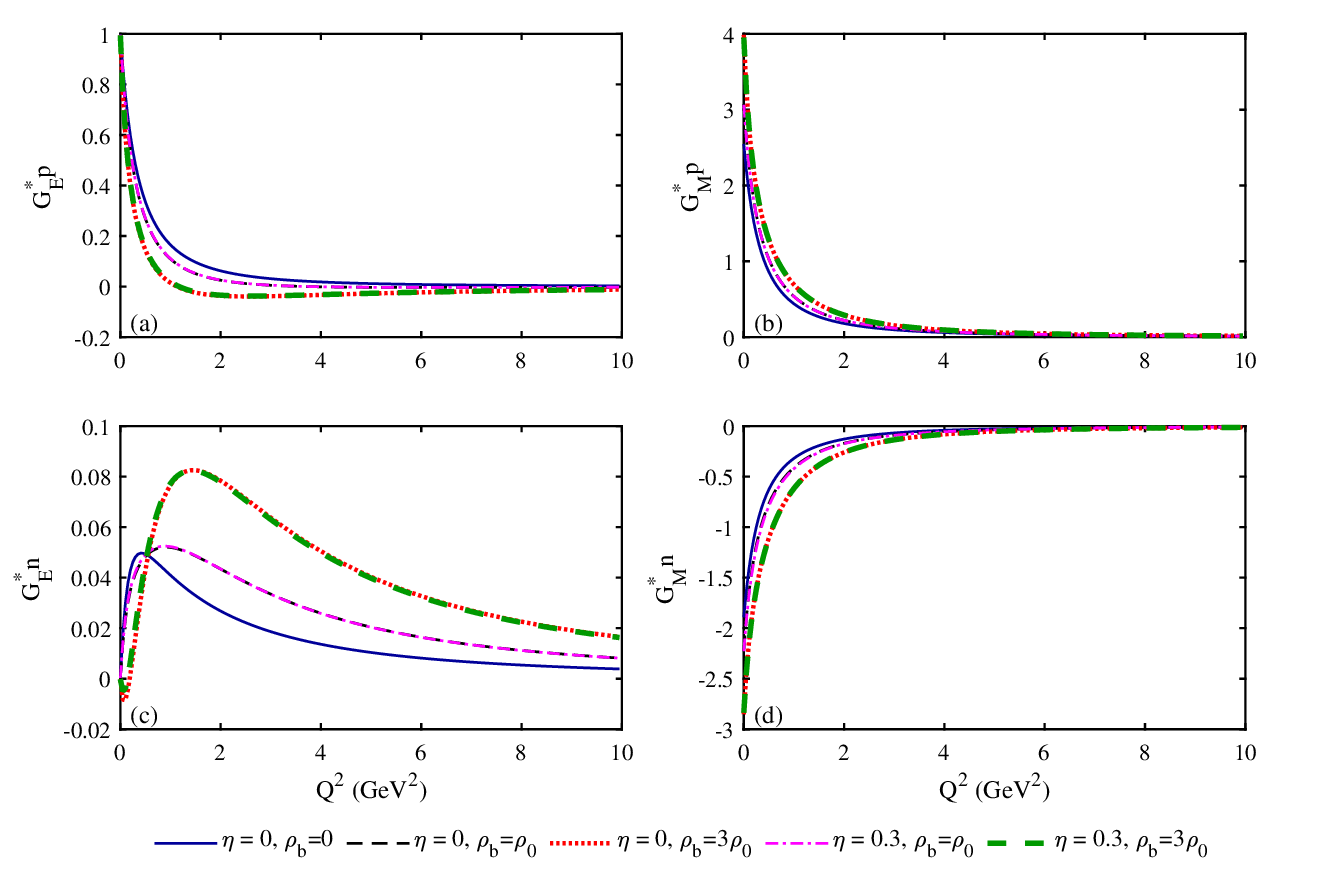}
		
			\caption{ The Sachs electric form factor $ G^{*p,n}_{E} $ (subplots(a) and (c)) and Sachs magnetic form factor $ G^{*p,n}_{M} $ (subplots(b) and (d)) for the baryons $p, n$, are shown as a function of momentum transfer $Q^2$ (in GeV$^2$) for different values of baryonic density $\rho_{B}$ and isospin asymmetry $\eta$ at temperature $T=0$ MeV.}
			\label{fig:figure1}
		
	\end{figure}
	\begin{figure}[h!]
	
		\includegraphics[height=1.0\textwidth,width=1.0\textwidth]{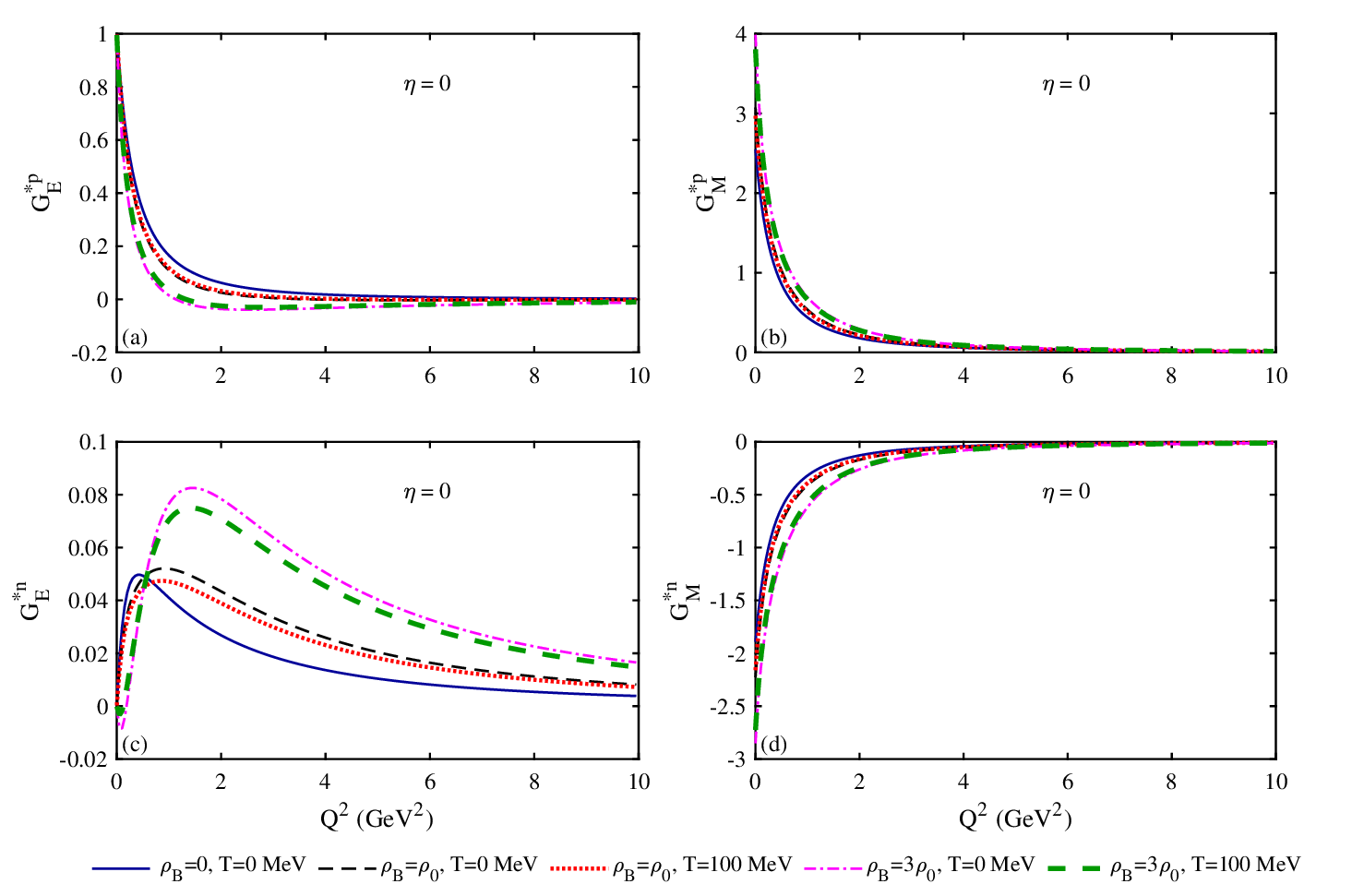}

	\caption{ Sachs electric form factor $ G^{*p,n}_{E} $ (subplots(a) and (c)) and Sachs magnetic form factor $ G^{*p,n}_{M} $ (subplots(b) and (d)) of the baryons $p, n$ are shown as a function of $Q^2$ (in GeV$^2$) in symmetric nuclear medium ($\eta = 0$), at temperatures $T= 0$ and $100$ MeV as well as baryonic densities $\rho_{B}=0, \rho_0$ and $3\rho$.}
	\label{fig:figure2}
	
\end{figure}
		\begin{figure}[h!]
		\includegraphics[height=.850\textwidth,width=1.1\textwidth]{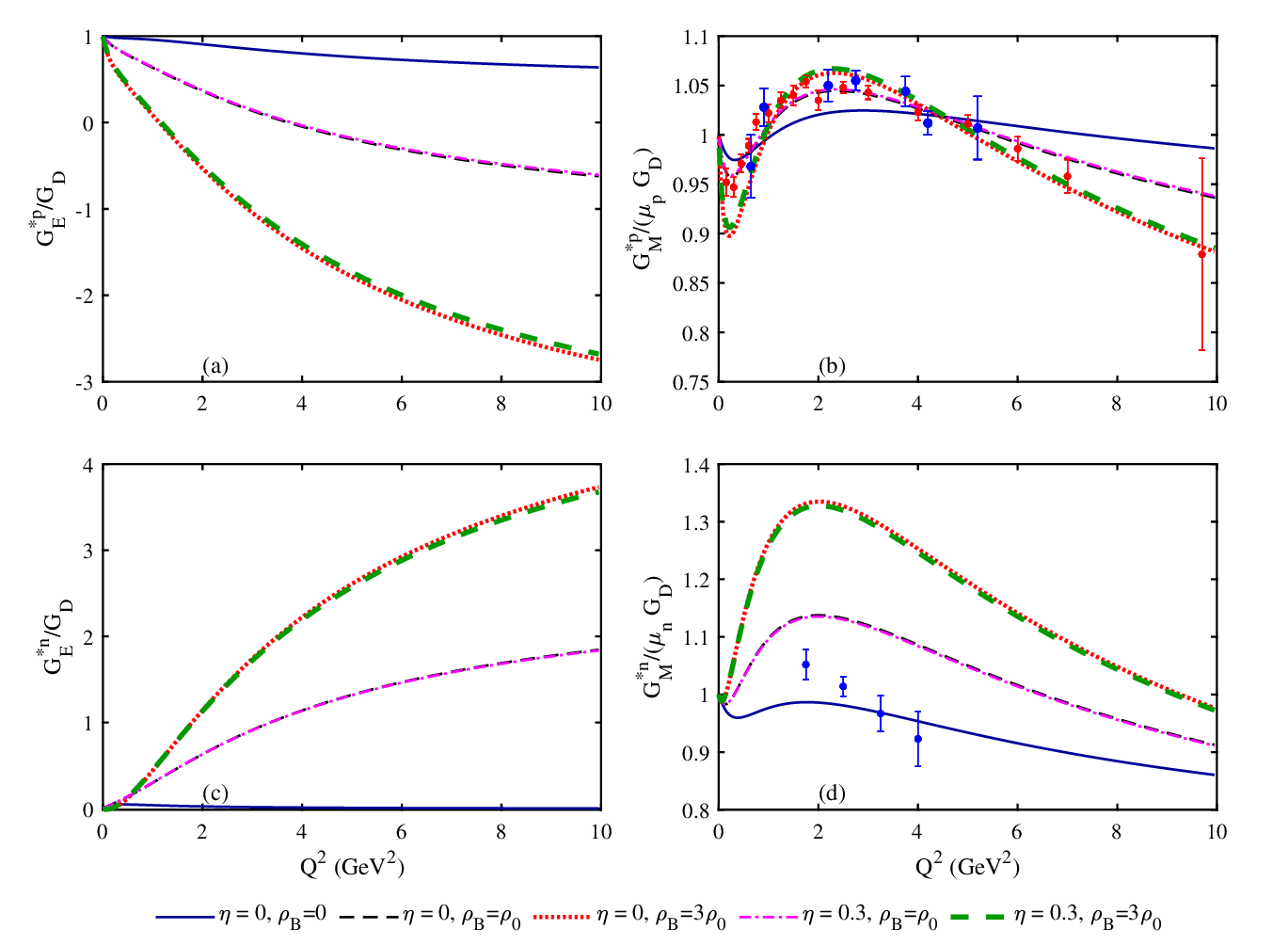}

		\caption{ Comparison of $ G^{*p,n}_{E} / G_D $ (subplots (a) and (c)) and $G^{*p,n}_{M} / (\mu^{*}_{p,n} G_D) $ (subplots (b) and (d)) for the baryons $p,n$ as a function of $Q^2$ (in GeV$^2$) for baryonic densities $\rho_B=0$, $\rho_0$ and $3\rho_0$ as well as isospin asymetry $\eta=0,0.3$ at temperature T$=0$ MeV. The experimental data for the case of proton shown by filled circles \cite{Walker:1993vj}, whereas the remaining proton data is taken from Ref.\cite{E94110:2004lsx}. The neutron magnetic form factor data is taken from Ref.\cite{Lung:1992bu}.}
		\label{fig:figure13}
		
	\end{figure}
	
				\begin{figure}[h!]
			\includegraphics[height=0.85\textwidth,width=1.10\textwidth]{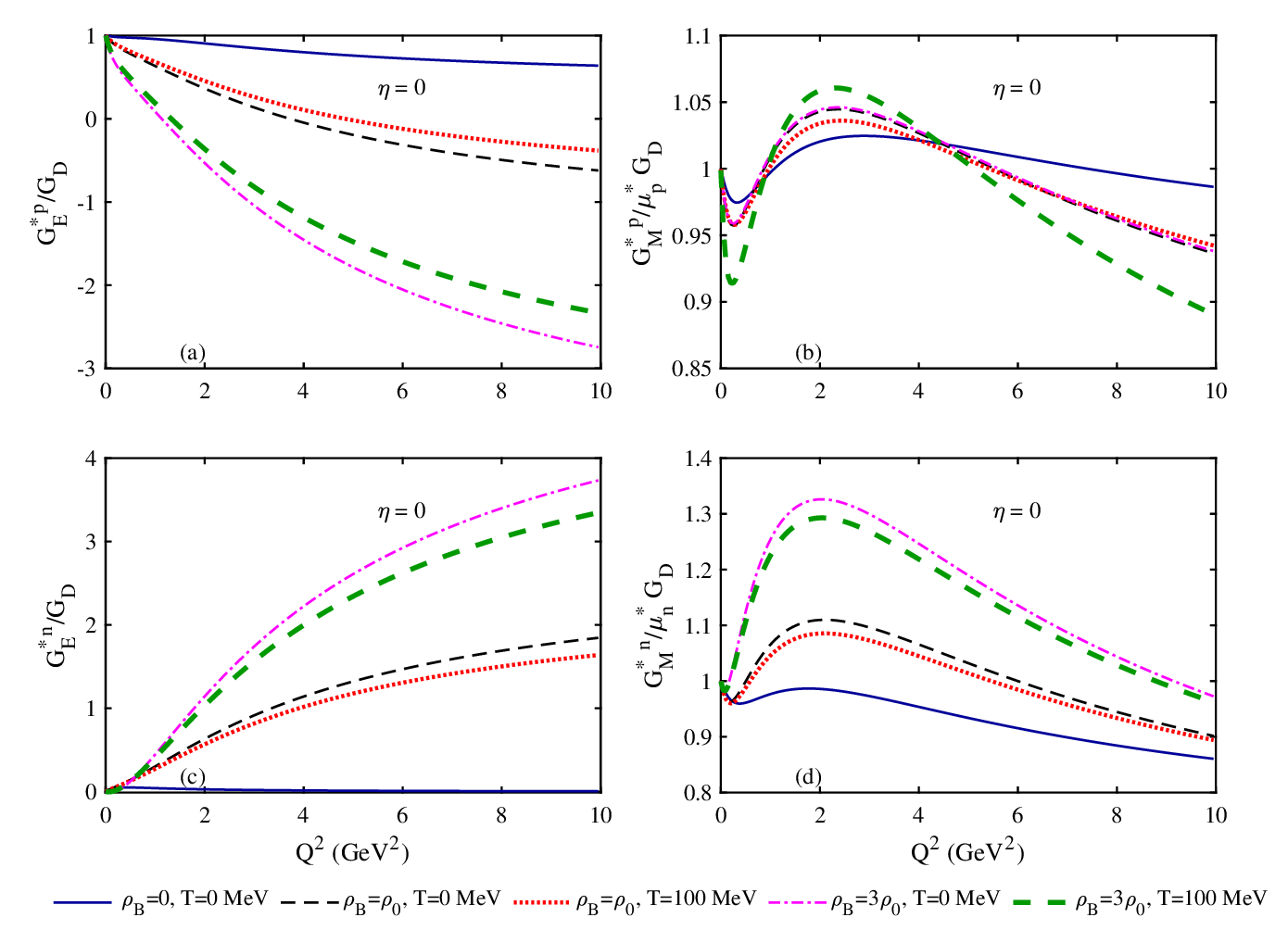}
			
				\caption{Comparison of the Sachs electric form factor ratio $ G^{*p,n}_{E} / G_D $ (subplots (a) and (c)) and Sachs magnetic form factor $G^{*p,n}_{M} / (\mu^{*}_{p,n} G_D) $ (subplots (b) and (d)) of the nucleon at $T=0, 100$ MeV and baryonic densities $\rho_{B}=0$ and $3\rho_0$.}
				\label{fig:figure3}
			\end{figure}
			
		\begin{figure}[h!]
			\includegraphics[height=0.85\textwidth,width=1.1\textwidth]{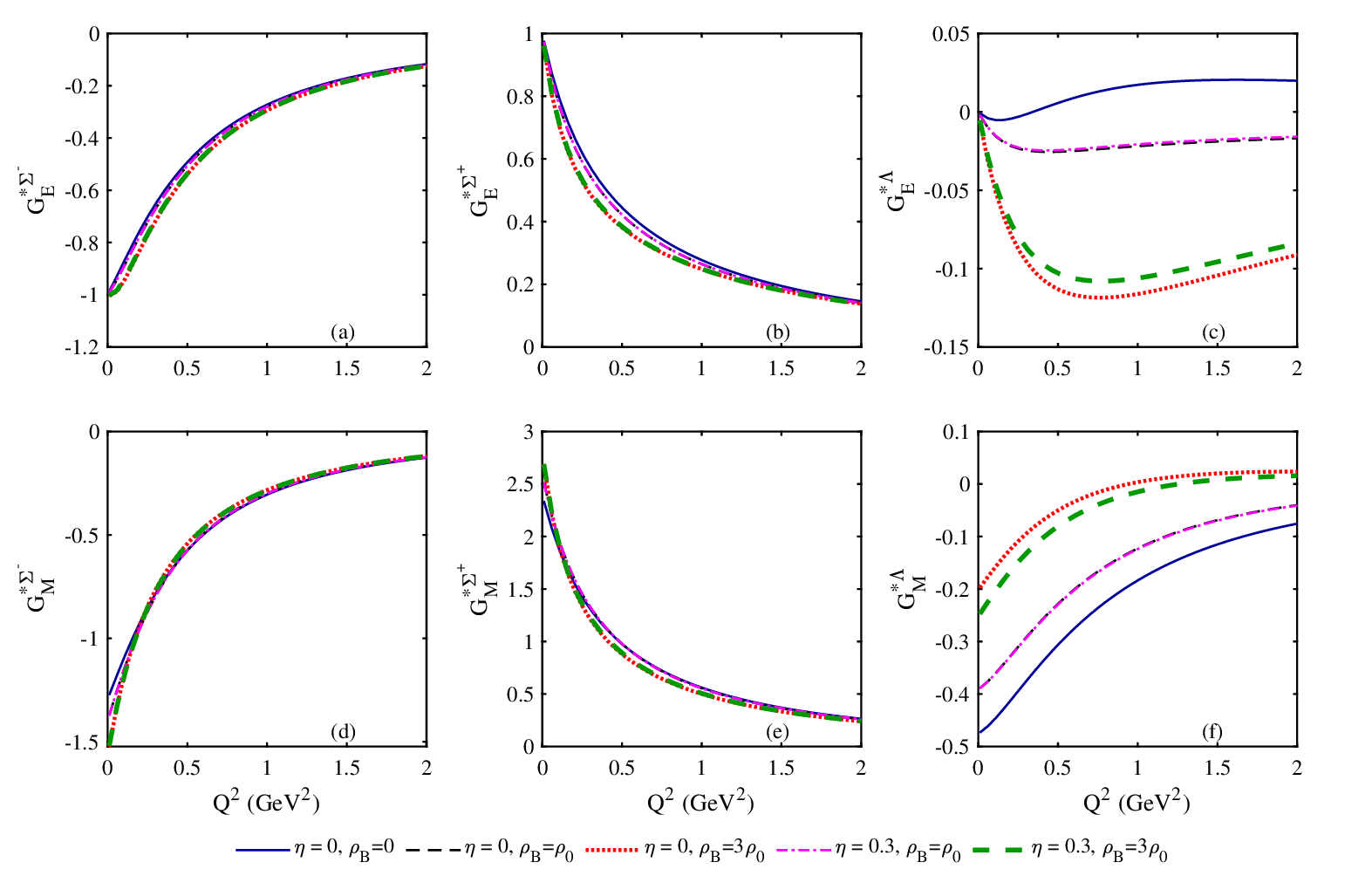}

			\caption{Comparison of the Sachs electric form factor $G^{*\Sigma^{\pm}, \Lambda}_{E}$ and Sachs magnetic form factor $G^{*\Sigma^{\pm}, \Lambda}_{M}$ of the $\Sigma^{\pm}$ and $\Lambda$ hyperon, are shown as a function of momentum transfer $Q^2$ (in GeV$^2$) at $T=0$ MeV.}
				\label{fig:figure4}
			\end{figure}	
			\begin{figure}[h!]
		\includegraphics[height=0.85\textwidth,width=1.1\textwidth]{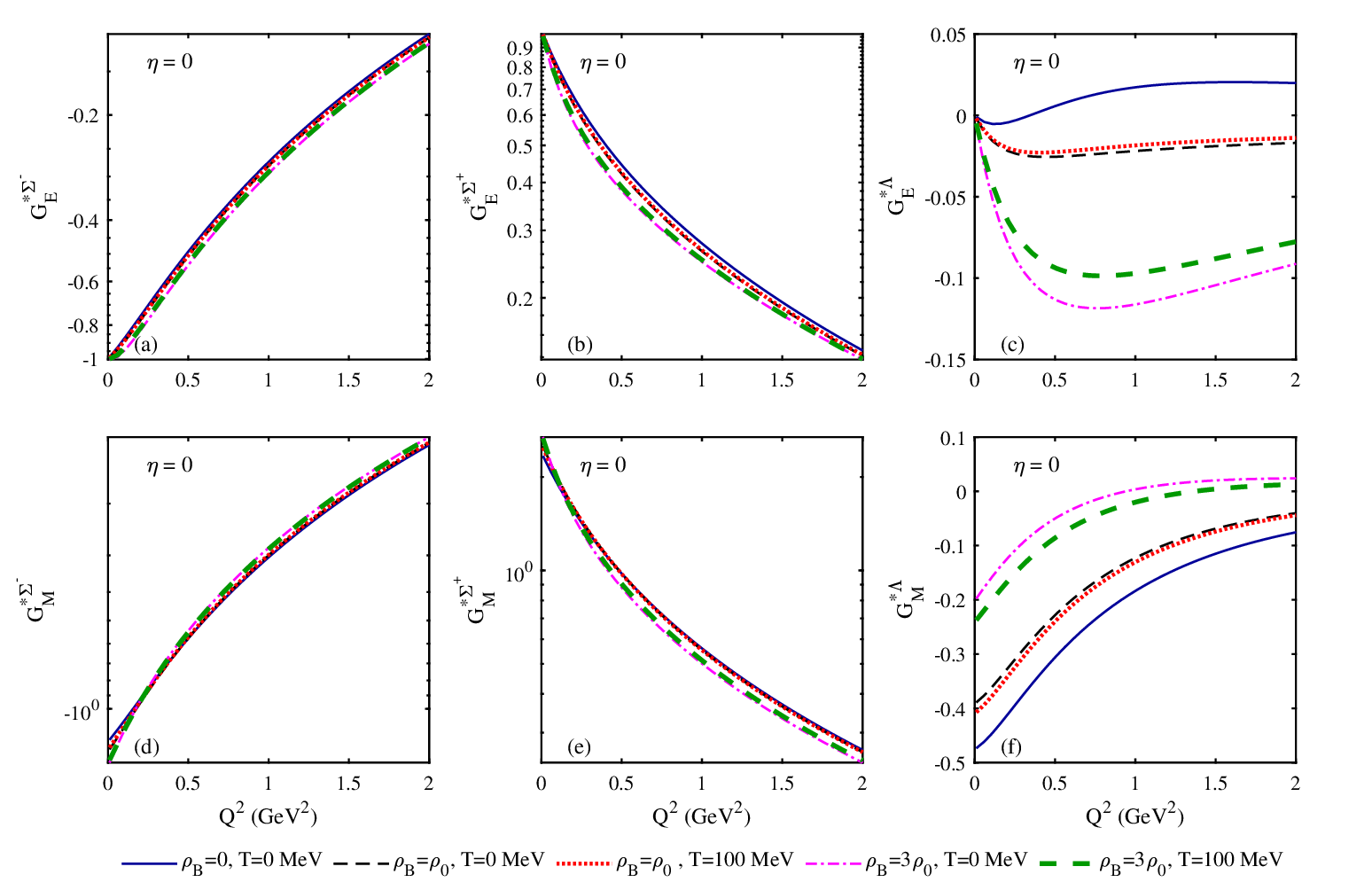}

			\caption{ Comparison of the Sachs electric form factor $G^{*\Sigma^{\pm}, \Lambda}_{E}$ and Sachs magnetic form factor $G^{*\Sigma^{\pm}, \Lambda}_{M}$ of the $\Sigma^{\pm}$ and $\Lambda$ hyperon shown as a function of momentum transfer $Q^2$ (in GeV$^2$) at $T=0, 100$ MeV and at different baryonic densities $\rho_{B}$.}
				\label{fig:figure5}
		\end{figure}

			\begin{figure}[h!]
		
				\includegraphics[height=0.85\textwidth,width=1.1\textwidth]{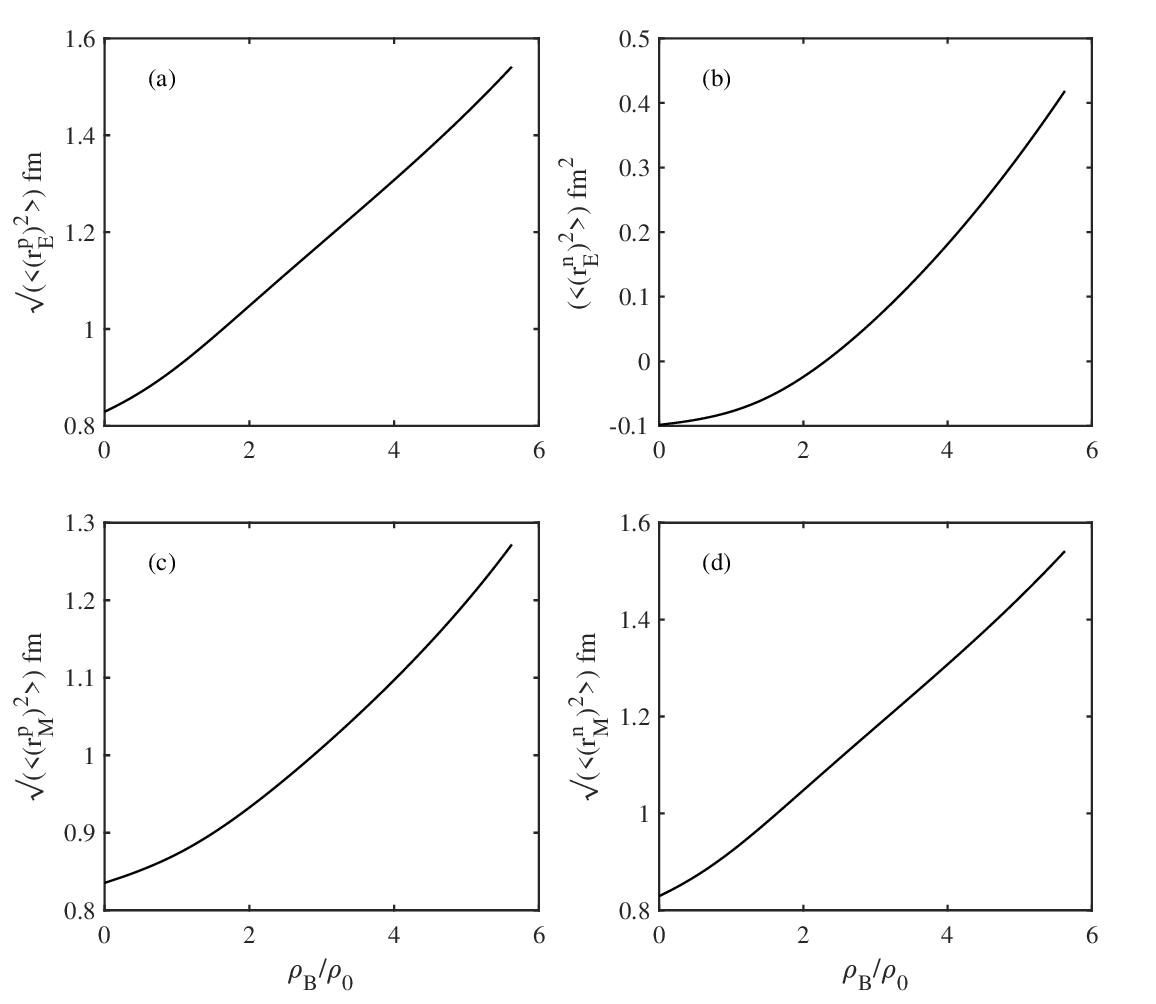}

			\caption{Electric charge radii 
				$\bigl\langle (r_E^{p})^2 \bigr\rangle^{1/2}$, 
				$\bigl\langle (r_E^{n})^2 \bigr\rangle^{}$ 
				and magnetic charge radii 
				$\bigl\langle (r_M^{p,n})^2 \bigr\rangle^{1/2}$ 
				of the nucleons as a function of baryon density $\rho_B$ (in units of $\rho_0$) at $T=0$ MeV and $Q^2=0$.}
				\label{fig:figure6}
		\end{figure}
				
		\begin{figure}[h!]
			\includegraphics[height=0.75\textwidth,width=1.1\textwidth]{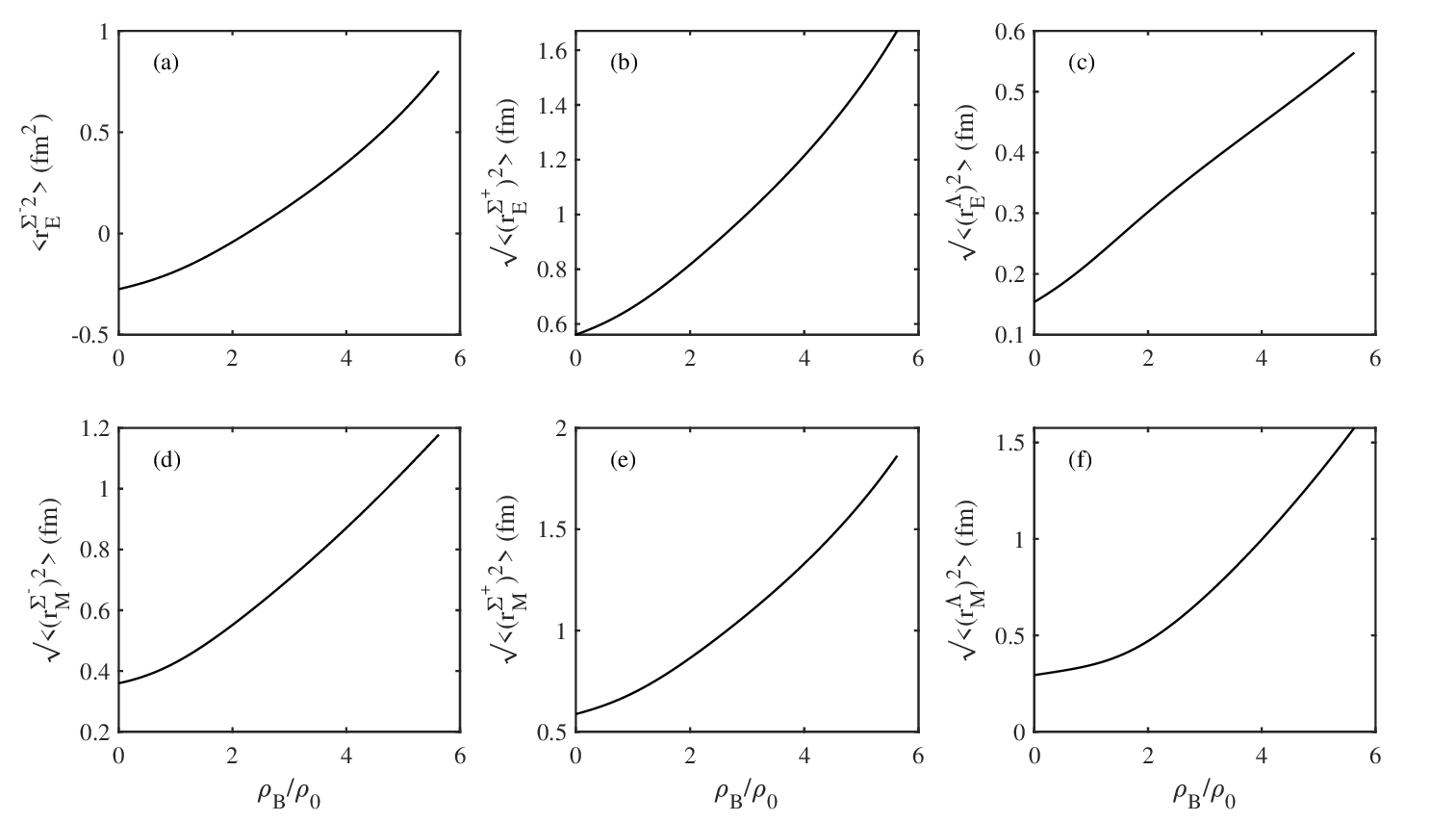}
			
		\caption{Electric charge radii 
			$\bigl\langle (r_E^{\Sigma^{+},\Lambda})^2 \bigr\rangle^{1/2}$, $\bigl\langle (r_E^{\Sigma^{-}})^2 \bigr\rangle^{}$  
			and magnetic charge radii 
			$\bigl\langle (r_M^{\Sigma^{\pm},\Lambda})^2 \bigr\rangle^{1/2}$ 
			of the hyperons as a function of baryon density $\rho_B$ (in units of $\rho_0$) at $T=0$ MeV and $Q^2=0$.}
				\label{fig:figure7}
		\end{figure}
		
\section{Results and Discussion}\label{sec:third}
\subsection{In-medium vector meson masses}

Using the CQMF model, we have minimised the thermodynamic potential $\Omega$ and obtained the in-medium values of the non-strange isoscalar field $\sigma$, the strange scalar field $\zeta$, scalar isovector field $\delta$ and dilaton field $\chi$. These field values have been used to calculate the quark and gluon condensates using Eqs.~\eqref{eq:eq2}--\eqref{eq:eq5}.
The scalar gluon condensates of QCD are simulated by a scalar dilaton field, $\chi$. In the present model, the values of the quark masses are taken as $m_u=4$ MeV, $m_d=7$ MeV and $m_s=150$ MeV. In vacuum, using Eqs.~\eqref{eq:eq2}--\eqref{eq:eq5} the value of condensate ($-m_q\langle \it{\bar{q}}q \rangle^{1/4}$) for $u, d$, and $s$-quark is found to be  $95.88$, $95.88$ and $248.16$ MeV, respectively using Eqs.~\eqref{eq:eq2}--\eqref{eq:eq5}. (In vacuum scalar isovector  field $\delta$ is $0$). All the three $u, d$ and $s$-condensates ($-m_q\langle \it{\bar{q}}q \rangle^{1/4}$) decrease with increase in baryonic density. On the other hand, with the increase in isospin asymmetry, the $u$-condensate increases and the $d$-condensate decreases which is clearly visible at higher baryonic density. For given isospin asymmetry and temperature of the nuclear matter, the scalar quark condensate is observed to decrease with increasing baryonic density.
  The NJL Model \cite{Klevansky:1992qe} describes the feature of the spontaneous chiral symmetry breaking of QCD which gives non zero values of the quark-antiquark condensates. In vacuum, the value of the scalar quark condensates ($\langle \it{\bar{q}}q \rangle^{}$) are obtained to be, $(-276.40$ MeV)$^3$, $(-229.51$ MeV)$^3$ and $(-293.5$ MeV)$^3$ for non-strange $u$ and $d$ and strange $s$-quark, respectively, with $m_u \neq m_d$. The same quantity as obtained from the NJL model are $(-247.8 $ MeV)$^3$, $(-247.8 $ MeV)$^3$ and  $(-258.3$ MeV)$^3$, respectively, with $m_u=m_d$ \cite{Klevansky:1992qe}.
 In the QCD sum rule, the in-medium masses of the light vector mesons have been calculated, and presented in Fig.~\ref{fig:fig0} for (a) $\rho$ meson ($m^*_\rho$), (b) $\omega$ meson ($m^*_\omega$), (c) $\phi$ meson ($m^*_\phi$) as a function of baryonic density $\rho_{B}$ at different values of temperature T and isospin asymmetry $\eta$. Using the finite energy sum rules in vacuum, the value of the parameters $k_u$, $k_d$ and $k_s$ appearing in Eqs.~\eqref{eq:eq13} and \eqref{eq:eq14}, are obtained as $3.59$, $5.64$ and $-1.13$, respectively.
For $\omega$ meson, at $T=0$ MeV, the effective mass obtained is $783.00$ MeV. At $T=100$ MeV, the effective mass is found to be $782.88$ MeV. The in-medium mass further increases with increase in isospin asymmetry. The effective masses decrease by $15.46\%$ and $41.69\%$ as baryonic density changes from $\rho_B=0$ to $\rho_0$ and $3\rho_0$, respectively. For the asymmetric medium $\eta=0.3$, the values of $m^*_\omega$ at $\rho_0$ and  $3\rho_0$  is observed to be $665.60$ and $472.31$ MeV, respectively. For $\rho$ meson, at vacuum, T$=0$ and T$=100$ MeV the effective masses are found to be $770.00$ and $769.88$ MeV, respectively. A visible effect of isospin asymmetry on $m^*_\rho$ has been observed at both $T=0$ MeV and $T=100$ MeV. For $\phi$ meson, at vacuum $T=0$ MeV, the effective mass is found to be $1020$ MeV. The in-medium mass of $\phi$ meson decreases with increase in baryonic density. A very slight variation with isospin asymmetry on the in-medium mass of the $\phi$ meson has been observed. The results obtained are consistent with the studies in Ref. \cite{Mishra:2014rha}.\\

\subsection{Nucleon's Sachs electric and magnetic form factors}
The variations of the nucleon's vacuum Sachs electric and Sachs magnetic form factors as function of $Q^2$ for $\eta=0$ and T$=0$ are shown in Fig.~\ref{fig:figure01} and are also compared with the available experimental data \cite{Lung:1992bu,Alexandrou:2018sjm}. The neutron's Sachs electric form factor results obtained aligns with the experimental data within error limits \cite{Schiavilla:2001qe}. The calculated Sachs form factors of proton and Sachs magnetic form factor of neutron within VMD framework are found to be in close agreement with lattice QCD results \cite{Alexandrou:2018sjm}. In Fig.~\ref{fig:figure02}, the vacuum Sachs form factors of the $\Sigma$ hyperons have been compared with the results obtained from lattice QCD and our results aligns with those in Ref. \cite{Liu:2023reo}.

The Sachs electric form factor of the proton $G^{*p}_E$ as a function of the $Q^2$ for different isospin asymmetry $\eta$ and baryonic density $\rho_{B}$ at T$=0$ MeV are shown in Fig.~\ref{fig:figure1} (a). The values of the $G^{*p}_E$ are observed to decrease monotonically with increasing $Q^2$ upto $Q^2=2-3$ and thereafter showing a saturation behavior. With increasing baryonic density $\rho_B$, the suppression of $G^{*p}_E $  is further noticeable, reflecting the enhanced medium effects at higher densities. With increasing isospin asymmetry, $\eta=0.3$ a negligible change in  $G^{*p}_E$ is seen. In Fig.~\ref{fig:figure1} (b), the proton Sachs magnetic form factor exhibits a rapid monotonic decrease in the low momentum transfer region $0 \leq Q^2 \, \leq 2$. In the intermediate region $2 \leq Q^2 \, \leq 4$, the decrease becomes weaker on the considered scale, while for higher momentum transfers 
$Q^2 \geq 4~$, the form factors show an almost constant behavior. An increase in the baryonic density from $\rho_B = 0$ to $3\rho_0$ leads to a systematic enhancement of the magnetic form factors over the entire range of $Q^2$ values. The presence of isospin asymmetry $\eta = 0.3$ has visible effect on these form factors. At finite temperature, only the magnitude of the form factors is noticeably affected, whereas the overall qualitative behavior remains unchanged.\\ In the Fig.~\ref{fig:figure1} (c), the Sachs electric form factor of the neutron, $G^{*n}_E $ at $T=0$  MeV shows a rapid rise in the low-$Q^2$ region and develops a pronounced maximum around $0.5 \leq Q^2 \leq 1.5$. The height of this peak shifts upwards with increase in baryonic density $\rho_{B}$, with an enhancement of approximately $10-20\%$ relative to the vacuum case. The peak shifts towards higher $Q^2$ with increasing baryonic density $\rho_{B}$ and negligible impact of isospin asymmetry $\eta$ is observed. Beyond the peak, $G^{*n}_E $ decreases monotonically with increasing $Q^2$, and values start converging at higher momentum transfer. The enhancement of $G^{*n}_E $ at low and intermediate values of $Q^2$ reflect the significant redistribution of the internal charge structure of the neutron in the dense matter whereas, Sachs magnetic form factor of the neutron $G^{*n}_M$, shows a monotonic increase with momentum transfer over the entire $Q^2$ range. $G^{*n}_M$ rises rapidly in the low-$Q^2$ region in the range $0 \leq Q^2 \leq 1.5~$. In the intermediate region, $1.5 \leq Q^2 \leq 4~$, the rate of increase becomes slower, and for higher momentum transfers $Q^2 \geq 4~$, $G^{*n}_M$ nearly saturates showing very weak $Q^2$ dependence. With increasing $\rho_B$ and isospin asymmetry, there is a suppression of the $G^{*n}_M$ over the entire $Q^2$ range, which has been depicted by the downward shift of the form factors.\\ The temperature dependence of Sachs form factors for nucleons have been presented in Fig.~\ref{fig:figure2}. At finite temperature $T=100$  MeV compared to $T=0$  MeV case, a further suppression of $G^{*p}_E $ has been observed as shown in Fig.~\ref{fig:figure2} (a), with the effect more pronounced at higher baryonic density $3\rho_0$. In proton's Sachs magnetic form factor temperature effect is slightly visible, mainly at low range of $Q^2$, as presented in  Fig. ~\ref{fig:figure2} (b). The value of $G_E(0)$ for both proton and neutron is the same in-medium and vacuum because the charge is preserved in the medium. The trends in Sachs form factors of neutron and  Sachs electric form factor of proton along with their variation with increasing $\rho_{B}$ is found to be consistent with the results obtained from the covariant quark model \cite{Ramalho:2025kii,Ramalho:2025vrm}. In the current study of Sachs magnetic form factor of the proton, $G^{*p}_M$ shows enhancement with increasing baryonic density $\rho_{B}$ and isospin asymmetry $\eta$. 
 For neutron’s Sachs electric form factor, $G_E^{*n}$ near the peak, a slight thermal enhancement is observed, and this effect becomes more visible at higher baryon density.  For $G_M^{*n}$ there is  again slight thermal enhancement with temperature, and more noticeable effect at larger baryonic densities in the low-$Q^2$ domain. Overall, a very small temperature dependence is visible at higher baryonic density on the form factors and their qualitative nature remains same.\\
  Fig.~\ref{fig:figure13} (a) shows the normalised in-medium proton's Sachs electric form factor at $T=0$ MeV, having monotonic decrease with increasing momentum transfer $Q^2$. The suppression in the value of the form factors increases with increasing in baryonic density $\rho_{B}$ over entire range of $Q^2$ values. The suppression is less at low momentum transfer $Q^2 \leq 1.5$ and becomes substantial at larger $Q^2$. However, the strongest suppression is observed at $3\rho_0$ indicating substantial modification of the proton electric structure in a dense nuclear medium. With increasing isospin asymmetry $\eta$, there is a negligible deviation in the low $Q^2$ range, whereas visible change at high $Q^2$ region particularly at the high baryonic densities. The curves exhibit a monotonically decreasing trend that steepens with increasing baryonic density. This represents physical ``swelling'' of the bound proton arising due to interactions with mean scalar fields. Similar impact of $\eta$ is also observed for $ G^{*p}_{M} /\mu_n G_D $ and $ G^{*n}_{E} / G_D $ ratios (Figs.~\ref{fig:figure13} (b) and (c), respectively).
 $G_M^{*p} / (\mu_p^* G_D)$ ratio in Fig.~\ref{fig:figure13} (b) depicts the effect of dense nuclear environment, modifying the meson cloud, thus enhancing the magnetic distribution relative to the dipole expectation. The $G_E^{*n} / G_D$ in Fig.~\ref{fig:figure13} (c) starts at zero and reveals how strong scalar and vector fields polarize the bound neutron by pulling internal charge distributions. The $G_M^{*n} / (\mu_n^* G_D)$ in Fig.~\ref{fig:figure13} (d) depicts higher sensitivity to the nuclear medium, signifying profound alterations in internal spin-flavor dynamics. The temperature dependence of the normalised in-medium nucleon's Sachs form factor's is shown in Fig.~\ref{fig:figure3}. The qualitative nature here remains the same with only a slight quantitative variation from the results obtained at $T=0$ MeV. 
 
 \subsection{$\Sigma{^\pm}$ and $\Lambda$ hyperon's EMFFs}
 
 Fig.~\ref{fig:figure4} shows Sachs form factors of the hyperons $G^{*\Sigma^{\pm}}_{E,M}$ and $G^{*\Lambda^{}}_{E,M}$ at $T=0$ MeV. The magnitude of $G^{*\Sigma^{+}}_{E}$, decreases monotonically with increasing momentum transfer over the entire range $0 \leq Q^2 \leq 2~$. Increase in baryonic density $\rho_{B}$ leads to the suppression of Sachs electric form factor. The increase in $\eta$ shows negligible impact as compared to the effect of $\rho_{B}$ on $G^{*\Sigma^{+}}_{E}$. Fig.~\ref{fig:figure4} (a) of $G^{*\Sigma^{-}}_{E}$, the Sachs electric form factor decreases in magnitude as a function of $Q^2$. These form factors suppresses with the increase in baryonic density. Increasing isospin asymmetry has negligible impact on these form factors.
 For example, as can be seen from Fig ~\ref{fig:figure4} (b), the maximum change occurs for 
 $Q^2 \leq 1~$ with suppression of about $10$-$15\%$ relative to the vacuum at $\rho_B=3\rho_0$. The in-medium effects decrease with increasing $Q^2$, whereas all curves tend to converge for 
 $Q^2 \approx 2$. Fig ~\ref{fig:figure4} (e) shows $G^{*\Sigma^{+}}_{M}$, which decreases monotonically with increasing $Q^2$. The Sachs magnetic form factor suppresses with increase in the baryonic density and has negligible impact of the isospin asymmetry. The magnetic form factors, $G^{*\Sigma^{-}}_{M}$ at $T=0$ MeV  are presented in Figs.~\ref{fig:figure4} (d) as a function of momentum transfer $Q^2$, with magnitude initially suppressed below $Q^2<0.2$ then increasing rapidly in the $Q^2$ region $0.2 \leq Q^2 \leq 1.5~$. The rate of increase is slower for $1.5 \leq Q^2 \leq 2~$. Baryons containing a larger number of strange quarks tend to be less affected by medium modifications and the results obtained in the present work are similar to those as observed in Ref. \cite{Ramalho:2025kii}. For $\Lambda$ hyperon as shown in Fig.~\ref{fig:figure4} (c), (f) the electric form factor $G^{*\Lambda^{}}_{E}$, and magnetic form factor $G^{*\Lambda^{}}_{M}$, respectively. The $G^{*\Lambda^{}}_{E}$ curves first decreases in magnitude, reaches a minima and then increases. The $G^{*\Lambda^{}}_{M}$, increases with increase in $Q^2$. There is visible change in form factors with increasing baryonic density and isospin asymmetry in both electric and magnetic form factors.\\ As shown in Fig.~\ref{fig:figure5}, Sachs electric and magnetic form factor of both $\Sigma^{\pm}$ shows very less temperature sensitivity. The results for the hyperon's Sachs electric and magnetic form factors as a function of $Q^2$ show a good agreement with the results obtained in Refs. \cite{Ramalho:2025kii,Ramalho:2012pu}. For the 
 neutral hyperon $\Lambda$, the temperature 
 dependence is relatively weaker for densities lower than 3$\rho_0$, with clear deviations at higher 
 densities as shown in Fig.~\ref{fig:figure5} (c) and comparatively less incase of Sachs magnetic  form factors in Fig.~\ref{fig:figure5} (f). Overall, these results indicate that the influence of temperature is amplified in the dense nuclear medium.\\ 
 
 \subsection{Charge Radii}
\hspace{-0.5cm} In the present study, we have calculated the electric and magnetic charge radii of the baryons, which can be respectively expressed as 
\begin{eqnarray*}
	\langle r^2 \rangle_E &=& -6 \left. \frac{dG_E(Q^2)}{dQ^2} \right|_{Q^2=0}~, \nonumber\\
	\quad \text{and} \quad\\
	\langle r^2 \rangle_M &=& -\frac{6}{\mu_b} \left. \frac{dG_M(Q^2)}{dQ^2} \right|_{Q^2=0}~.
	\label{eq:eq39}
\end{eqnarray*}
 The values of electric and magnetic charge radii obtained from the current study have been summarised in Table ~\ref{tab:table4}, and are compared with the experimental results \cite{Xiong:2019umf,Atac:2021wqj}. The variations of the electric and magnetic charge radii as a function of baryonic density are presented in Figs.~\ref{fig:figure6} and~\ref{fig:figure7}, respectively. We have observed that the  electric and magnetic charge radii for the baryons increases with baryonic density as witnessed through the in-medium modification of the form factors. Increasing baryonic densities lead to redistribution of electric charge and magnetisation in the nucleons and thus changing the internal structure in the dense nuclear matter.
Fig.~\ref{fig:figure6} shows the electric and magnetic charge radii for the case of proton and neutron. The charge radii are observed to increase with an increase in the baryonic density. For proton, the electric charge radius at $\rho_B=0$ is $0.829$ fm, which changes to $0.922$ fm at and at $\rho_B=\rho_0$ and $1.186$ fm at $3\rho_0$. For proton the magnetic charge radii value at $\rho_B=0$ is found to be $0.832$ fm, which changes to $0.876$ and $1.015$ fm at  $\rho_B=\rho_0$ and $3\rho_0$, respectively. For neutron, enhancement in the charge radii with baryonic density has been observed and is given in Table ~\ref{tab:table4}.\\
 Fig.~\ref{fig:figure7} shows the electric and magnetic charge radii of the $\Sigma^{\pm}$ and $\Lambda$ hyperon. For $\Sigma^+$, the electric charge radius value at $\rho_B=0$ is $0.561$ fm and changes to $0.6616$ and $1.0023$ fm at $\rho_B=\rho_0$ and $3\rho_0$, respectively. The magnetic charge radius of $\Sigma^+$ at $\rho_B=0$ is $0.587$ fm and changes to $0.691$ and $1.079$ fm at $\rho_B=\rho_0$ and $3\rho_0$, respectively. An increase has been observed in the charge radii with the baryonic density for $\Sigma^-$ and $\Lambda$  hyperon and are presented in Table ~\ref{tab:table4}.
The vacuum values obtained are consistent with the experimental and other theoretical models\cite{Simon:1980hu, Bijker:2005cd, Hyde:2004gef,Kopecky:1997rw,Kubon:2001rj,Yang:2019mzq,Xiong:2019umf}.
 
\section{Summary}\label{sec:fourth}

To summarize, in the current study, the quark and gluon condensates have been investigated in the CQMF model which are further taken as inputs to calculate the effective masses of the light vector mesons using the QCD sum rule approach. The dependence of masses of the light vector mesons on the baryonic density has been studied and the masses are found to decrease with increase in baryonic density. The introduction of isospin asymmetry in the medium is observed to cause splitting in the corresponding quarks and quark condensates and hence in the in-medium meson masses. Isospin asymmetry effects are observed to be more dominant for the mesons composed of the light $u$ and $d$ quarks, whereas strangeness effect become less noticeable for the mesons having strange quarks due to its weak interaction with the nuclear matter.\\ The Dirac and Pauli form factors of the baryons calculated within the framework of VMD model are dependent on the effective masses of the vector mesons under consideration which are then used to calculate the Sachs form factors, dipole ratios and charge radii of the baryons. The Sachs form factors shows an appreciable dependence on baryonic density, but slight dependence on temperature and isospin asymmetry. The modification of Sachs form factors in the nuclear matter varies from one baryon to another and depend on the baryon's properties differently. The charge radii at vacuum are found close to the experimental values. The electric and magnetic charge radii of the baryons are found to increase with increasing baryonic density.\\
 The calculations at higher baryonic densities are important as they can be used for studies with the astrophysical implications. Higher density (upto $\rho_B$=3$\rho_0$) represents neutron-star core density where possible exotic phases and dense matter effects become significant. The present study of the in-medium form factors of the baryons will be useful for the future experimental facilities like EIC. The current work can be extended to obtain the density dependent EMFFs in the time-like region.

\section*{Acknowledgement}\label{sec:fifth}

\hspace{-0.5cm} H.D. would like to thank the Science and Engineering Research Board, Anusandhan-National Research Foundation, Government of India under the scheme SERB-POWER Fellowship (Ref No. SPF/2023/000116) for financial support. A. K. sincerely acknowledges Anushandhan National Research Foundation, Government of India for funding of the research project under the Science and Engineering Research Board Core Research Grant (SERB-CRG) scheme (File No. CRG/2023/000557).
\clearpage
\hspace*{0.95cm}

\end{document}